\documentclass[journal,10pt,twoside,twocolumn]{IEEEtran}

\usepackage{cite}
\usepackage[T1]{fontenc}
\usepackage{graphicx}

\usepackage{amssymb}
\usepackage{amsmath}
\usepackage{paralist}
\usepackage{subfigure}

\usepackage{amssymb}
\usepackage{amsfonts}
\usepackage{mathrsfs}
\usepackage{xspace}
\usepackage{bm}
\usepackage{upgreek}

\newcommand{\safemath}[2]{\newcommand{#1}{\ensuremath{#2}\xspace}}

\safemath{\bma}{\mathbf{a}}
\safemath{\bmb}{\mathbf{b}}
\safemath{\bmc}{\mathbf{c}}
\safemath{\bmd}{\mathbf{d}}
\safemath{\bme}{\mathbf{e}}
\safemath{\bmf}{\mathbf{f}}
\safemath{\bmg}{\mathbf{g}}
\safemath{\bmh}{\mathbf{h}}
\safemath{\bmi}{\mathbf{i}}
\safemath{\bmj}{\mathbf{j}}
\safemath{\bmk}{\mathbf{k}}
\safemath{\bml}{\mathbf{l}}
\safemath{\bmm}{\mathbf{m}}
\safemath{\bmn}{\mathbf{n}}
\safemath{\bmo}{\mathbf{o}}
\safemath{\bmp}{\mathbf{p}}
\safemath{\bmq}{\mathbf{q}}
\safemath{\bmr}{\mathbf{r}}
\safemath{\bms}{\mathbf{s}}
\safemath{\bmt}{\mathbf{t}}
\safemath{\bmu}{\mathbf{u}}
\safemath{\bmv}{\mathbf{v}}
\safemath{\bmw}{\mathbf{w}}
\safemath{\bmx}{\mathbf{x}}
\safemath{\bmy}{\mathbf{y}}
\safemath{\bmz}{\mathbf{z}}
\safemath{\bmzero}{\mathbf{0}}
\safemath{\bmone}{\mathbf{1}}

\bmdefine{\biad}{a}
\bmdefine{\bibd}{b}
\bmdefine{\bicd}{c}
\bmdefine{\bidd}{d}
\bmdefine{\bied}{e}
\bmdefine{\bifd}{f}
\bmdefine{\bigd}{g}
\bmdefine{\bihd}{h}
\bmdefine{\biid}{i}
\bmdefine{\bijd}{j}
\bmdefine{\bikd}{k}
\bmdefine{\bild}{l}
\bmdefine{\bimd}{m}
\bmdefine{\bind}{n}
\bmdefine{\biod}{o}
\bmdefine{\bipd}{p}
\bmdefine{\biqd}{q}
\bmdefine{\bird}{r}
\bmdefine{\bisd}{s}
\bmdefine{\bitd}{t}
\bmdefine{\biud}{u}
\bmdefine{\bivd}{v}
\bmdefine{\biwd}{w}
\bmdefine{\bixd}{x}
\bmdefine{\biyd}{y}
\bmdefine{\bizd}{z}

\bmdefine{\bixid}{\xi}
\bmdefine{\bilambdad}{\lambda}
\bmdefine{\bimud}{\mu}
\bmdefine{\bithetad}{\theta}
\bmdefine{\biphid}{\phi}
\bmdefine{\bideltad}{\delta}

\safemath{\bmia}{\biad}
\safemath{\bmib}{\bibd}
\safemath{\bmic}{\bicd}
\safemath{\bmid}{\bidd}
\safemath{\bmie}{\bied}
\safemath{\bmif}{\bifd}
\safemath{\bmig}{\bigd}
\safemath{\bmih}{\bihd}
\safemath{\bmii}{\biid}
\safemath{\bmij}{\bijd}
\safemath{\bmik}{\bikd}
\safemath{\bmil}{\bild}
\safemath{\bmim}{\bimd}
\safemath{\bmin}{\bind}
\safemath{\bmio}{\biod}
\safemath{\bmip}{\bipd}
\safemath{\bmiq}{\biqd}
\safemath{\bmir}{\bird}
\safemath{\bmis}{\bisd}
\safemath{\bmit}{\bitd}
\safemath{\bmiu}{\biud}
\safemath{\bmiv}{\bivd}
\safemath{\bmiw}{\biwd}
\safemath{\bmix}{\bixd}
\safemath{\bmiy}{\biyd}
\safemath{\bmiz}{\bizd}

\safemath{\bmxi}{\bixid}
\safemath{\bmlambda}{\bilambdad}
\safemath{\bmmu}{\bimud}
\safemath{\bmtheta}{\bithetad}
\safemath{\bmphi}{\biphid}
\safemath{\bmdelta}{\bideltad}

\safemath{\bA}{\mathbf{A}}
\safemath{\bB}{\mathbf{B}}
\safemath{\bC}{\mathbf{C}}
\safemath{\bD}{\mathbf{D}}
\safemath{\bE}{\mathbf{E}}
\safemath{\bF}{\mathbf{F}}
\safemath{\bG}{\mathbf{G}}
\safemath{\bH}{\mathbf{H}}
\safemath{\bI}{\mathbf{I}}
\safemath{\bJ}{\mathbf{J}}
\safemath{\bK}{\mathbf{K}}
\safemath{\bL}{\mathbf{L}}
\safemath{\bM}{\mathbf{M}}
\safemath{\bN}{\mathbf{N}}
\safemath{\bO}{\mathbf{O}}
\safemath{\bP}{\mathbf{P}}
\safemath{\bQ}{\mathbf{Q}}
\safemath{\bR}{\mathbf{R}}
\safemath{\bS}{\mathbf{S}}
\safemath{\bT}{\mathbf{T}}
\safemath{\bU}{\mathbf{U}}
\safemath{\bV}{\mathbf{V}}
\safemath{\bW}{\mathbf{W}}
\safemath{\bX}{\mathbf{X}}
\safemath{\bY}{\mathbf{Y}}
\safemath{\bZ}{\mathbf{Z}}

\safemath{\bZero}{\mathbf{0}}
\safemath{\bOne}{\mathbf{1}}
\safemath{\bDelta}{\mathbf{\Delta}}
\safemath{\bLambda}{\mathbf{\UpLambda}}
\safemath{\bPhi}{\mathbf{\Upphi}}
\safemath{\bSigma}{\mathbf{\Upsigma}}
\safemath{\bOmega}{\mathbf{\Upomega}}
\safemath{\bTheta}{\mathbf{\Uptheta}}

\bmdefine{\biAd}{A}
\bmdefine{\biBd}{B}
\bmdefine{\biCd}{C}
\bmdefine{\biDd}{D}
\bmdefine{\biEd}{E}
\bmdefine{\biFd}{F}
\bmdefine{\biGd}{G}
\bmdefine{\biHd}{H}
\bmdefine{\biId}{I}
\bmdefine{\biJd}{J}
\bmdefine{\biKd}{K}
\bmdefine{\biLd}{L}
\bmdefine{\biMd}{M}
\bmdefine{\biOd}{N}
\bmdefine{\biPd}{O}
\bmdefine{\biQd}{P}
\bmdefine{\biRd}{R}
\bmdefine{\biSd}{S}
\bmdefine{\biTd}{T}
\bmdefine{\biUd}{U}
\bmdefine{\biVd}{V}
\bmdefine{\biWd}{W}
\bmdefine{\biXd}{X}
\bmdefine{\biYd}{Y}
\bmdefine{\biZd}{Z}

\bmdefine{\biDelta}{\Delta}
\bmdefine{\biLambda}{\Lambda}
\bmdefine{\biPhi}{\Phi}
\bmdefine{\biSigma}{\Sigma}
\bmdefine{\biOmega}{\Omega}
\bmdefine{\biTheta}{\Theta}

\safemath{\bimA}{\biAd}
\safemath{\bimB}{\biBd}
\safemath{\bimC}{\biCd}
\safemath{\bimD}{\biDd}
\safemath{\bimE}{\biEd}
\safemath{\bimF}{\biFd}
\safemath{\bimG}{\biGd}
\safemath{\bimH}{\biHd}
\safemath{\bimI}{\biId}
\safemath{\bimJ}{\biJd}
\safemath{\bimK}{\biKd}
\safemath{\bimL}{\biLd}
\safemath{\bimM}{\biMd}
\safemath{\bimN}{\biNd}
\safemath{\bimO}{\biOd}
\safemath{\bimP}{\biPd}
\safemath{\bimQ}{\biQd}
\safemath{\bimR}{\biRd}
\safemath{\bimS}{\biSd}
\safemath{\bimT}{\biTd}
\safemath{\bimU}{\biUd}
\safemath{\bimV}{\biVd}
\safemath{\bimW}{\biWd}
\safemath{\bimX}{\biXd}
\safemath{\bimY}{\biYd}
\safemath{\bimZ}{\biZd}

\safemath{\bimDelta}{\biDelta}
\safemath{\bimLambda}{\biLambda}
\safemath{\bimPhi}{\biPhi}
\safemath{\bimSigma}{\biSigma}
\safemath{\bimOmega}{\biOmega}
\safemath{\bimTheta}{\biTheta}

\safemath{\setA}{\mathcal{A}}
\safemath{\setB}{\mathcal{B}}
\safemath{\setC}{\mathcal{C}}
\safemath{\setD}{\mathcal{D}}
\safemath{\setE}{\mathcal{E}}
\safemath{\setF}{\mathcal{F}}
\safemath{\setG}{\mathcal{G}}
\safemath{\setH}{\mathcal{H}}
\safemath{\setI}{\mathcal{I}}
\safemath{\setJ}{\mathcal{J}}
\safemath{\setK}{\mathcal{K}}
\safemath{\setL}{\mathcal{L}}
\safemath{\setM}{\mathcal{M}}
\safemath{\setN}{\mathcal{N}}
\safemath{\setO}{\mathcal{O}}
\safemath{\setP}{\mathcal{P}}
\safemath{\setQ}{\mathcal{Q}}
\safemath{\setR}{\mathcal{R}}
\safemath{\setS}{\mathcal{S}}
\safemath{\setT}{\mathcal{T}}
\safemath{\setU}{\mathcal{U}}
\safemath{\setV}{\mathcal{V}}
\safemath{\setW}{\mathcal{W}}
\safemath{\setX}{\mathcal{X}}
\safemath{\setY}{\mathcal{Y}}
\safemath{\setZ}{\mathcal{Z}}
\safemath{\emptySet}{\varnothing}

\safemath{\colA}{\mathscr{A}}
\safemath{\colB}{\mathscr{B}}
\safemath{\colC}{\mathscr{C}}
\safemath{\colD}{\mathscr{D}}
\safemath{\colE}{\mathscr{E}}
\safemath{\colF}{\mathscr{F}}
\safemath{\colG}{\mathscr{G}}
\safemath{\colH}{\mathscr{H}}
\safemath{\colI}{\mathscr{I}}
\safemath{\colJ}{\mathscr{J}}
\safemath{\colK}{\mathscr{K}}
\safemath{\colL}{\mathscr{L}}
\safemath{\colM}{\mathscr{M}}
\safemath{\colN}{\mathscr{N}}
\safemath{\colO}{\mathscr{O}}
\safemath{\colP}{\mathscr{P}}
\safemath{\colQ}{\mathscr{Q}}
\safemath{\colR}{\mathscr{R}}
\safemath{\colS}{\mathscr{S}}
\safemath{\colT}{\mathscr{T}}
\safemath{\colU}{\mathscr{U}}
\safemath{\colV}{\mathscr{V}}
\safemath{\colW}{\mathscr{W}}
\safemath{\colX}{\mathscr{X}}
\safemath{\colY}{\mathscr{Y}}
\safemath{\colZ}{\mathscr{Z}}

\safemath{\opA}{\mathbb{A}}
\safemath{\opB}{\mathbb{B}}
\safemath{\opC}{\mathbb{C}}
\safemath{\opD}{\mathbb{D}}
\safemath{\opE}{\mathbb{E}}
\safemath{\opF}{\mathbb{F}}
\safemath{\opG}{\mathbb{G}}
\safemath{\opH}{\mathbb{H}}
\safemath{\opI}{\mathbb{I}}
\safemath{\opJ}{\mathbb{J}}
\safemath{\opK}{\mathbb{K}}
\safemath{\opL}{\mathbb{L}}
\safemath{\opM}{\mathbb{M}}
\safemath{\opN}{\mathbb{N}}
\safemath{\opO}{\mathbb{O}}
\safemath{\opP}{\mathbb{P}}
\safemath{\opQ}{\mathbb{Q}}
\safemath{\opR}{\mathbb{R}}
\safemath{\opS}{\mathbb{S}}
\safemath{\opT}{\mathbb{T}}
\safemath{\opU}{\mathbb{U}}
\safemath{\opV}{\mathbb{V}}
\safemath{\opW}{\mathbb{W}}
\safemath{\opX}{\mathbb{X}}
\safemath{\opY}{\mathbb{Y}}
\safemath{\opZ}{\mathbb{Z}}
\safemath{\opZero}{\mathbb{O}}
\safemath{\identityop}{\opI}

\safemath{\veca}{\bma}
\safemath{\vecb}{\bmb}
\safemath{\vecc}{\bmc}
\safemath{\vecd}{\bmd}
\safemath{\vece}{\bme}
\safemath{\vecf}{\bmf}
\safemath{\vecg}{\bmg}
\safemath{\vech}{\bmh}
\safemath{\veci}{\bmi}
\safemath{\vecj}{\bmj}
\safemath{\veck}{\bmk}
\safemath{\vecl}{\bml}
\safemath{\vecm}{\bmm}
\safemath{\vecn}{\bmn}
\safemath{\veco}{\bmo}
\safemath{\vecp}{\bmp}
\safemath{\vecq}{\bmq}
\safemath{\vecr}{\bmr}
\safemath{\vecs}{\bms}
\safemath{\vect}{\bmt}
\safemath{\vecu}{\bmu}
\safemath{\vecv}{\bmv}
\safemath{\vecw}{\bmw}
\safemath{\vecx}{\bmx}
\safemath{\vecy}{\bmy}
\safemath{\vecz}{\bmz}

\safemath{\veczero}{\bmzero}
\safemath{\vecone}{\bmone}
\safemath{\vecxi}{\bmxi}
\safemath{\veclambda}{\bmlambda}
\safemath{\vecmu}{\bmmu}
\safemath{\vectheta}{\bmtheta}
\safemath{\vecphi}{\bmphi}
\safemath{\vecdelta}{\bmdelta}

\safemath{\matA}{\bA}
\safemath{\matB}{\bB}
\safemath{\matC}{\bC}
\safemath{\matD}{\bD}
\safemath{\matE}{\bE}
\safemath{\matF}{\bF}
\safemath{\matG}{\bG}
\safemath{\matH}{\bH}
\safemath{\matI}{\bI}
\safemath{\matJ}{\bJ}
\safemath{\matK}{\bK}
\safemath{\matL}{\bL}
\safemath{\matM}{\bM}
\safemath{\matN}{\bN}
\safemath{\matO}{\bO}
\safemath{\matP}{\bP}
\safemath{\matQ}{\bQ}
\safemath{\matR}{\bR}
\safemath{\matS}{\bS}
\safemath{\matT}{\bT}
\safemath{\matU}{\bU}
\safemath{\matV}{\bV}
\safemath{\matW}{\bW}
\safemath{\matX}{\bX}
\safemath{\matY}{\bY}
\safemath{\matZ}{\bZ}
\safemath{\matzero}{\bmzero}

\safemath{\matDelta}{\bDelta}
\safemath{\matLambda}{\bLambda}
\safemath{\matPhi}{\bPhi}
\safemath{\matSigma}{\bSigma}
\safemath{\matOmega}{\bOmega}
\safemath{\matTheta}{\bTheta}

\safemath{\matidentity}{\matI}
\safemath{\matone}{\matO}

\safemath{\rnda}{A}
\safemath{\rndb}{B}
\safemath{\rndc}{C}
\safemath{\rndd}{D}
\safemath{\rnde}{E}
\safemath{\rndf}{F}
\safemath{\rndg}{G}
\safemath{\rndh}{H}
\safemath{\rndi}{I}
\safemath{\rndj}{J}
\safemath{\rndk}{K}
\safemath{\rndl}{L}
\safemath{\rndm}{M}
\safemath{\rndn}{N}
\safemath{\rndo}{O}
\safemath{\rndp}{P}
\safemath{\rndq}{Q}
\safemath{\rndr}{R}
\safemath{\rnds}{S}
\safemath{\rndt}{T}
\safemath{\rndu}{U}
\safemath{\rndv}{V}
\safemath{\rndw}{W}
\safemath{\rndx}{X}
\safemath{\rndy}{Y}
\safemath{\rndz}{Z}

\safemath{\rveca}{\bimA}
\safemath{\rvecb}{\bimB}
\safemath{\rvecc}{\bimC}
\safemath{\rvecd}{\bimD}
\safemath{\rvece}{\bimE}
\safemath{\rvecf}{\bimF}
\safemath{\rvecg}{\bimG}
\safemath{\rvech}{\bimH}
\safemath{\rveci}{\bimI}
\safemath{\rvecj}{\bimJ}
\safemath{\rveck}{\bimK}
\safemath{\rvecl}{\bimL}
\safemath{\rvecm}{\bimM}
\safemath{\rvecn}{\bimN}
\safemath{\rveco}{\bomO}
\safemath{\rvecp}{\bimP}
\safemath{\rvecq}{\bimQ}
\safemath{\rvecr}{\bimR}
\safemath{\rvecs}{\bimS}
\safemath{\rvect}{\bimT}
\safemath{\rvecu}{\bimU}
\safemath{\rvecv}{\bimV}
\safemath{\rvecw}{\bimW}
\safemath{\rvecx}{\bimX}
\safemath{\rvecy}{\bimY}
\safemath{\rvecz}{\bimZ}

\safemath{\rvecxi}{\bmxi}
\safemath{\rveclambda}{\bmlambda}
\safemath{\rvecmu}{\bmmu}
\safemath{\rvectheta}{\bmtheta}
\safemath{\rvecphi}{\bmphi}

\safemath{\rmatA}{\bimA}
\safemath{\rmatB}{\bimB}
\safemath{\rmatC}{\bimC}
\safemath{\rmatD}{\bimD}
\safemath{\rmatE}{\bimE}
\safemath{\rmatF}{\bimF}
\safemath{\rmatG}{\bimG}
\safemath{\rmatH}{\bimH}
\safemath{\rmatI}{\bimI}
\safemath{\rmatJ}{\bimJ}
\safemath{\rmatK}{\bimK}
\safemath{\rmatL}{\bimL}
\safemath{\rmatM}{\bimM}
\safemath{\rmatN}{\bimN}
\safemath{\rmatO}{\bimO}
\safemath{\rmatP}{\bimP}
\safemath{\rmatQ}{\bimQ}
\safemath{\rmatR}{\bimR}
\safemath{\rmatS}{\bimS}
\safemath{\rmatT}{\bimT}
\safemath{\rmatU}{\bimU}
\safemath{\rmatV}{\bimV}
\safemath{\rmatW}{\bimW}
\safemath{\rmatX}{\bimX}
\safemath{\rmatY}{\bimY}
\safemath{\rmatZ}{\bimZ}

\safemath{\rmatDelta}{\bimDelta}
\safemath{\rmatLambda}{\bimLambda}
\safemath{\rmatPhi}{\bimPhi}
\safemath{\rmatSigma}{\bimSigma}
\safemath{\rmatOmega}{\bimOmega}
\safemath{\rmatTheta}{\bimTheta}

\usepackage{amssymb}
\usepackage{amsfonts}
\usepackage{mathrsfs}
\usepackage{xspace}
\usepackage{bm}
\usepackage{fancyref}
\usepackage{textcomp}

\usepackage{multirow}
\usepackage{stmaryrd}

\newenvironment{textbmatrix}{	\setlength{\arraycolsep}{2.5pt}%
								\big[\begin{matrix}}{\end{matrix}\big]%
								\raisebox{0.08ex}{\vphantom{M}}}

\def\be{\begin{equation}}
\def\ee{\end{equation}}
\def\een{\nonumber \end{equation}}
\def\mat{\begin{bmatrix}}
\def\emat{\end{bmatrix}}
\def\btm{\begin{textbmatrix}}
\def\etm{\end{textbmatrix}}

\def\ba#1\ea{\begin{align}#1\end{align}}
\def\bas#1\eas{\begin{align*}#1\end{align*}}
\def\bs#1\es{\begin{split}#1\end{split}} 
\def\bg#1\eg{\begin{gather}#1\end{gather}}
\def\bml#1\eml{\begin{multline}#1\end{multline}}
\def\bi#1\ei{\begin{itemize}#1\end{itemize}}

\newcommand{\lefto}{\mathopen{}\left}

\DeclareMathOperator*{\argmin}{arg\;min}		
\DeclareMathOperator*{\argmax}{arg\;max}		

\newcommand{\abs}[1]{\lefto\lvert#1\right\rvert}		

\newcommand{\vecnorm}[1]{\lefto\lVert#1\right\rVert}		
\newcommand{\herm}[1]{\ensuremath{#1^{H}}} 	

\safemath{\dirac}{\delta}					
\safemath{\krond}{\dirac}					
\newcommand{\allonotwo}[2]{\ensuremath{#1=1,\ldots,#2}}

\safemath{\upto}{\uparrow}
\safemath{\downto}{\downarrow}
\safemath{\iu}{j}							
\safemath{\ev}{\lambda}						
\safemath{\hilseqspace}{l^{2}}				
\newcommand{\banachfunspace}[1]{\setL^{#1}}	
\safemath{\hilfunspace}{\banachfunspace{2}}	

\safemath{\SNR}{\text{\sc snr}} 				
\safemath{\No}{N_0}							
\safemath{\Es}{E_s}							
\safemath{\Eb}{E_b}							
\safemath{\EbNo}{\frac{\Eb}{\No}}
\safemath{\EsNo}{\frac{\Es}{\No}}

\DeclareMathOperator{\CHop}{\ensuremath{\opH}} 
\safemath{\tvir}{\rndh_{\CHop}}				
\safemath{\tvtf}{\rndl_{\CHop}}				
\safemath{\spf}{\rnds_{\CHop}}				
\safemath{\bff}{H_{\CHop}}					

\safemath{\ircf}{r_{h}}						
\safemath{\tftvcf}{r_{s}}					
\safemath{\tfcf}{r_{l}}						
\safemath{\bfcf}{r_{H}}						

\safemath{\tcorr}{c_h}						
\safemath{\scf}{c_{s}}						
\safemath{\tfcorr}{c_{l}}					
\safemath{\fcorr}{c_{H}}						

\safemath{\mi}{I}							
\safemath{\capacity}{C}						

\safemath{\normal}{\mathcal{N}}			
\safemath{\jpg}{\mathcal{CN}}			
\safemath{\mchain}{\leftrightarrow}		

\safemath{\dB}{\,\mathrm{dB}}
\safemath{\dBm}{\,\mathrm{dBm}}
\safemath{\Hz}{\,\mathrm{Hz}}
\safemath{\kHz}{\,\mathrm{kHz}}
\safemath{\MHz}{\,\mathrm{MHz}}
\safemath{\GHz}{\,\mathrm{GHz}}
\safemath{\s}{\,\mathrm{s}}
\safemath{\ms}{\,\mathrm{ms}}
\safemath{\mus}{\,\mathrm{\text{\textmu}s}}
\safemath{\ns}{\,\mathrm{ns}}
\safemath{\ps}{\,\mathrm{ps}}
\safemath{\meter}{\,\mathrm{m}}
\safemath{\mm}{\,\mathrm{mm}}
\safemath{\cm}{\,\mathrm{cm}}
\safemath{\m}{\,\mathrm{m}}
\safemath{\W}{\,\mathrm{W}}
\safemath{\mW}{\, \mathrm{mW}}
\safemath{\J}{\,\mathrm{J}}
\safemath{\K}{\,\mathrm{K}}
\safemath{\bit}{\,\mathrm{bit}}
\safemath{\nat}{\,\mathrm{nat}}

\safemath{\define}{\triangleq}			

\safemath{\equivalent}{\sim}
\safemath{\distas}{\sim}					
\safemath{\sdiff}{\Delta}				

\safemath{\reals}{\mathbb{R}}
\safemath{\positivereals}{\reals_{+}}
\safemath{\integers}{\mathbb{Z}}
\safemath{\posint}{\integers_{+}}
\safemath{\naturals}{\mathbb{N}}
\safemath{\posnaturals}{\naturals_{+}}
\safemath{\complexset}{\mathbb{C}}
\safemath{\rationals}{\mathbb{Q}}

\newcommand*{\fancyrefapplabelprefix}{app}		
\newcommand*{\fancyrefthmlabelprefix}{thm}		
\newcommand*{\fancyreflemlabelprefix}{lem}		
\newcommand*{\fancyrefcorlabelprefix}{cor}		
\newcommand*{\fancyrefdeflabelprefix}{def}		
\newcommand*{\fancyrefproplabelprefix}{prop}		
\newcommand*{\fancyrefexmpllabelprefix}{exmpl}
\frefformat{vario}{\fancyrefseclabelprefix}{Section~#1}
\frefformat{vario}{\fancyrefthmlabelprefix}{Theorem~#1}
\frefformat{vario}{\fancyreflemlabelprefix}{Lemma~#1}
\frefformat{vario}{\fancyrefcorlabelprefix}{Corollary~#1}
\frefformat{vario}{\fancyrefdeflabelprefix}{Definition~#1}
\frefformat{vario}{\fancyreffiglabelprefix}{Fig.~#1}
\frefformat{vario}{\fancyrefapplabelprefix}{Appendix~#1} 
\frefformat{vario}{\fancyrefeqlabelprefix}{(#1)}
\frefformat{vario}{\fancyrefproplabelprefix}{Property~#1}
\frefformat{vario}{\fancyrefexmpllabelprefix}{Example~#1}

 \newcommand{\trm}{\textrm}
 \newtheorem{thm}{Theorem}
 \newtheorem{cor}[thm]{Corollary}   
 
 \newtheorem{defi}{Definition}

\safemath{\dictab}{[\,\dicta\,\,\dictb\,]}

\safemath{\ysig}{\bmy}
\safemath{\ysighat}{\hat{\ysig}}
\safemath{\ysigdim}{M}
\safemath{\xsig}{\bmx}
\safemath{\xsigdim}{N}
\safemath{\nx}{n_x}
\safemath{\zsig}{\bmz}
\safemath{\zsigdim}{\ysigdim}
\safemath{\rsig}{\bmr}
\safemath{\Adict}{\bA}
\safemath{\Adicttilde}{\widetilde{\Adict}}
\safemath{\Adictdim}{\outputdim\times\xsigdim}
\safemath{\avec}{\bma}
\safemath{\avectilde}{\tilde{\avec}}
\safemath{\Bdict}{\bB}
\safemath{\Bdicttilde}{\widetilde{\Bdict}}
\safemath{\Cdict}{\bC}
\safemath{\cvec}{\bmc}
\safemath{\Ddict}{\bD}
\safemath{\Ddictdim}{\ysigdim\times\xsigdim}
\safemath{\dvec}{\bmd}
\safemath{\Ddicttilde}{\widetilde{\bD}}
\safemath{\Bonb}{\bB}
\safemath{\bvec}{\bmb}
\safemath{\Bonbdim}{\ysigdim\times\ysigdim}
\safemath{\noise}{\bmn}
\safemath{\noisedim}{\ysigim}
\safemath{\err}{\bme}
\safemath{\errdim}{\ysigdim}
\safemath{\errset}{\setE}
\safemath{\nerr}{n_e}
\safemath{\delop}{\bP_\errset}
\safemath{\delopc}{\bP_{{\errset}^c}}

\safemath{\cplxi}{\imath}
\safemath{\cplxj}{\jmath}
\newcommand{\comb}[1]{\vecdelta_{#1}}
\safemath{\dict}{\matD}
\safemath{\inputdim}{N}		
\safemath{\outputdim}{M}		
\safemath{\sparsity}{S}	
\safemath{\inputdimA}{{N_a}}	
\safemath{\inputdimB}{{N_b}}	
\safemath{\elemA}{{n_a}}	
\safemath{\elemB}{{n_b}}	
\safemath{\resA}{\matR_a}	
\safemath{\resB}{\matR_b}	
\safemath{\subD}{\matS} 
\safemath{\subA}{\matS_a} 
\safemath{\subB}{\matS_b} 
\safemath{\dicta}{\matA} 	
\safemath{\dictb}{\matB} 	
\safemath{\hollowS}{H}
\safemath{\hollowA}{H_a}
\safemath{\hollowB}{H_b}
\safemath{\cross}{Z}
\safemath{\coh}{\mu_d}			
\safemath{\coha}{\mu_a}			
\safemath{\cohb}{\mu_b}			
\safemath{\mubs}{\nu}	
\safemath{\cohm}{\mu_m} 
\safemath{\dictset}{\setD}	
\safemath{\dictsetp}{\dictset(\coh,\coha,\cohb)}	
\safemath{\dictsetgen}{\dictset_\text{gen}}
\safemath{\dictsetgenp}{\dictsetgen(\coh)}
\safemath{\dictsetonb}{\dictset_\text{onb}}
\safemath{\dictsetonbp}{\dictsetonb(\coh)}

\safemath{\leftside}{U}
\safemath{\rightsideA}{R_a}
\safemath{\rightsideB}{R_b}

\safemath{\indexS}{\setI_S} 

\safemath{\na}{n_a}			
\safemath{\nb}{n_b}			
\safemath{\coeffa}{p_i}	
\safemath{\coeffb}{q_j}	
\safemath{\seta}{\setP}		
\safemath{\setb}{\setQ}     
\safemath{\setw}{\setW}	
\safemath{\setz}{\setZ}	
\safemath{\cola}{\veca}		
\safemath{\colb}{\vecb}		
\safemath{\cold}{\vecd}		
\safemath{\inputvec}{\vecx} 	
\safemath{\error}{\vece}	
\safemath{\noiseout}{\vecz} 	
\safemath{\inputvecel}{x}
\safemath{\inputveca}{\vecx_a}
\safemath{\inputvecb}{\vecx_b}
\safemath{\outputvec}{\vecy}	
\safemath{\lambdamin}{\lambda_{\mathrm{min}}}

\newcommand{\pos}[1]{\lefto[#1\right]^+}
\newcommand{\normtwo}[1]{\vecnorm{#1}_2}
\newcommand{\normone}[1]{\vecnorm{#1}_1}
\newcommand{\normzero}[1]{\vecnorm{#1}_0}

\safemath{\elltwo}{\ell_2}
\safemath{\ellone}{\ell_1}
\safemath{\ellzero}{\ell_0}
\safemath{\ellinf}{\ell_\infty}
\safemath{\licard}{Z(\coh,\coha,\cohb)}
\safemath{\xsol}{\hat{x}}
\safemath{\xbord}{x_b}		
\safemath{\xstat}{x_s}		
\safemath{\xstatLone}{\tilde{x}_s}
\safemath{\order}{\mathcal{O}} 
\safemath{\scales}{\Theta} 
\safemath{\ones}{\mathbf{1}} 
\safemath{\zeroes}{\mathbf{0}} 
\safemath{\thlone}{\kappa(\coh,\cohb)} 
\safemath{\constoneA}{\delta} 
\safemath{\constoneB}{\epsilon} 
\safemath{\nlarge}{L}				   
\safemath{\sumlarge}{S_\nlarge}
\safemath{\maxlarger}{P_\nlarge}	   
\safemath{\Pzero}{\textrm{P0}}	
\safemath{\Pone}{\textrm{P1}}
\safemath{\vecfir}{\vecw}			 
\safemath{\vecsec}{\vecz}
\safemath{\elvecfir}{w}              
\safemath{\elvecsec}{z}				 
\safemath{\nlargefir}{n}
\safemath{\normout}{\gamma}
\safemath{\auxfun}{h}
\safemath{\supp}{\textrm{supp}}

\safemath{\indexa}{\ell}
\safemath{\indexb}{r}
\safemath{\indexc}{i}
\safemath{\indexd}{j}

\safemath{\project}{P}

\usepackage{color}
\newcommand{\revision}[1]{#1}

\newcounter{MYtempeqncnt}

\begin{document}

\title{Recovery of Sparsely Corrupted Signals}

\author{Christoph~Studer,~\IEEEmembership{Member,~IEEE}, Patrick Kuppinger,~\IEEEmembership{Student Member,~IEEE}, \\ Graeme Pope,~\IEEEmembership{Student Member,~IEEE}, and Helmut~B\"olcskei,~\IEEEmembership{Fellow,~IEEE}

  \thanks{Part of this paper was {presented at} the IEEE International Symposium on Information Theory (ISIT), Saint-Petersburg, Russia, July 2011~\cite{SKPB11c}. \revision{This work was supported in part by the Swiss National Science Foundation (SNSF) under Grant~PA00P2-134155.}}
 
  \thanks{\revision{C.~Studer was with the Dept.~of Information Technology and Electrical Engineering, ETH Zurich, Switzerland, and is now with the Dept.~of Electrical and Computer Engineering, Rice University, Houston, TX, USA (e-mail: studer@rice.edu).}}
  \thanks{\revision{P.~Kuppinger was with the Dept.~of Information Technology and Electrical Engineering, ETH Zurich, Switzerland, and is now with UBS, Zurich, Switzerland (e-mail: patrick.kuppinger@gmail.com).}}
  \thanks{\revision{G.~Pope and H.~B\"olcskei are with the Dept.~of Information Technology and Electrical Engineering, ETH Zurich, Switzerland (e-mail: gpope@nari.ee.ethz.ch; boelcskei@nari.ee.ethz.ch).}}


}

\markboth{To appear in IEEE Transactions on Information Theory}{Studer \emph{et al.}: Recovery of Sparsely Corrupted Signals}

\IEEEpubid{XXXXX.00~\copyright~2012 IEEE}

\maketitle


\begin{abstract}
We investigate the recovery of signals exhibiting a sparse representation in a general (i.e., possibly redundant or incomplete) dictionary that are corrupted by additive noise admitting a sparse representation in another general dictionary. 
This setup covers a wide range of applications, such as image inpainting, super-resolution, signal separation, and recovery of signals that are impaired by, e.g., clipping, impulse noise, or narrowband interference.
We present deterministic recovery guarantees based on a novel uncertainty relation for pairs of general dictionaries and we provide corresponding practicable recovery algorithms.
The recovery guarantees we find depend on the signal and noise sparsity levels, on the coherence parameters of the involved dictionaries, and on the amount of prior knowledge about the signal and noise support sets.
%
\end{abstract}




\section{Introduction}\label{sec:intro}
We consider the problem of identifying the sparse vector~$\inputvec\in\complexset^\inputdimA$ from~$\outputdim$ linear and non-adaptive measurements collected in the vector
\begin{align} \label{eq:systemmodel}
 \noiseout = \dicta \inputvec + \dictb\error
\end{align}
where  $\dicta\in\complexset^{\outputdim\times \inputdimA}$ and $\dictb\in\complexset^{\outputdim\times \inputdimB}$ 
are known deterministic and general (i.e., not necessarily of the same cardinality, and possibly redundant or incomplete) \emph{dictionaries}, and $\error\in\complexset^{\inputdimB}$ represents a sparse noise vector. 
The support set of \error and the corresponding nonzero entries can be \emph{arbitrary}; in particular, \error may also depend on~$\inputvec$ and/or the dictionary~$\dicta$.

%
%

This recovery problem occurs in many applications, some of which are described next:
\begin{itemize}

\item \emph{Clipping:} \revision{Non-linearities in (power-)amplifiers or in analog-to-digital converters often cause signal clipping or saturation~\cite{abel1991}. 
This impairment can be cast into the signal model \fref{eq:systemmodel} by setting $\dictb=\bI_\outputdim$, where $\bI_\outputdim$ denotes the $\outputdim\times\outputdim$ identity matrix, and rewriting \fref{eq:systemmodel} as $\noiseout=\outputvec + \error$ with $\error = g_a(\outputvec) - \outputvec$.
Concretely, instead of the $M$-dimensional signal vector $\outputvec=\dicta\inputvec$ of interest, the device in question delivers $g_a(\outputvec)$, where the function $g_a(\outputvec)$ realizes entry-wise signal clipping to the interval $[-a,+a]$.  
The vector $\error$ will be sparse, provided  the clipping level is high enough. 
Furthermore, in this case the support set of \error can be identified prior to recovery, by simply comparing the absolute values of the entries of \outputvec to the clipping threshold~$a$. 
%
%
Finally, we note that here it is essential that the noise vector \error  be allowed to depend on the vector \inputvec and/or the dictionary~\dicta.}


\IEEEpubidadjcol

\item \emph{Impulse noise:} In numerous applications, one has to deal with the recovery of signals corrupted by impulse noise~\cite{carrillo2010}. Specific applications include, e.g., reading out from unreliable memory~\cite{novak2010} or recovery of audio signals impaired by click/pop noise, which typically occurs during playback of old phonograph records. 
The model in~\fref{eq:systemmodel} is easily seen to incorporate such impairments.
Just set $\dictb=\bI_\outputdim$ and let \error be the impulse-noise vector.
We would like to emphasize the generality of~\fref{eq:systemmodel} which allows impulse noise that is sparse in general dictionaries~\dictb.

\item \emph{Narrowband interference:} In many applications one is interested in recovering audio, video, or communication signals that are corrupted by narrowband interference. Electric hum, as it may occur in improperly designed audio or video equipment, is a typical example of such an impairment.
Electric hum typically exhibits a sparse representation in the Fourier basis as it (mainly) consists of a tone at some base-frequency and a series of corresponding harmonics, which is captured by setting $\dictb=\bF_\outputdim$ in~\fref{eq:systemmodel}, \revision{where $\bF_\outputdim$ is the $\outputdim$-dimensional discrete Fourier transform (DFT) matrix defined below in \fref{eq:fouriermatrix}.}


\item \emph{Super-resolution and inpainting:} Our framework also encompasses super-resolution~\cite{mallat2010,elad2001} and inpainting~\cite{bertalmio2000} for images, audio, and video signals.
In both applications, only a subset of the entries of the (full-resolution) signal vector $\outputvec=\dicta\inputvec$ is available and the task is to fill in the missing entries of the signal vector such that $\outputvec=\dicta\inputvec$. 
The missing entries are accounted for by choosing the vector \error such that the entries of $\noiseout=\outputvec+\error$ corresponding to the missing entries in \outputvec are set to some (arbitrary) value, e.g., $0$.
The missing entries of~\outputvec are then filled in by first recovering \inputvec from \noiseout and then computing $\outputvec=\dicta\inputvec$. 
%
%
%
%
%
Note that in both applications the support set \setE is known (i.e., the locations of the missing entries can easily be identified) and the dictionary \dicta is typically redundant (see, e.g., \cite{aharon2006} for a corresponding discussion), i.e., \dicta has more dictionary elements (columns) than rows, which demonstrates the need for recovery results that apply to general (i.e., possibly redundant) dictionaries.

\item \emph{Signal separation:} Separation of (audio or video) signals into two distinct components also fits into our framework.
%
A prominent example for this task is the separation of texture from cartoon parts in images (see~\cite{elad2005,donoho2010} and references therein). 
In the language of our setup, the dictionaries \dicta and \dictb are chosen such that they allow for sparse representation of the two distinct features; \inputvec and \error are the corresponding coefficients describing these features (sparsely). 
Note that here the vector \error no longer plays the role of (undesired) noise.
Signal separation then amounts to simultaneously extracting the sparse vectors \inputvec and \error from the observation (e.g., the image) $\noiseout = \dicta\inputvec + \dictb\error$.
%

\end{itemize}

Naturally, it is of significant practical interest to identify fundamental limits on the recovery of \inputvec (and \error, if appropriate) from \noiseout in~\fref{eq:systemmodel}. 
For the noiseless case $\noiseout=\dicta\inputvec$ such \emph{recovery guarantees} are known~\cite{donoho2002,gribonval2003,tropp2004} and typically set limits on the maximum allowed number of nonzero entries of \inputvec or---more colloquially---on the ``sparsity'' level of \inputvec. 
%
%
\revision{These recovery guarantees are usually expressed in terms of restricted isometry constants (RICs)\cite{candes2005decoding,C08} or in terms of the coherence parameter \cite{donoho2002,gribonval2003,tropp2004,CWX10} of the dictionary \dicta. 
In contrast to coherence parameters, RICs can, in general, not be computed efficiently. In this paper, we focus exclusively on coherence-based recovery guarantees.} 
%
 %
For the case of unstructured noise, i.e., $\noiseout=\dicta\inputvec + \noise$ with no constraints imposed on \noise apart from $\normtwo{\noise}<\infty$, coherence-based recovery guarantees were derived in~\cite{donoho2006a,fuchs2006,tropp2006,benhaim2010,CWX10}.
The corresponding results, however, do not guarantee perfect recovery of \inputvec, but only ensure that either the recovery error is bounded above by a function of $\normtwo{\noise}$ or only guarantee perfect recovery of the support set of \inputvec.
Such results are to be expected, as a consequence of the generality of the setup in terms of the assumptions on the noise vector~\vecn.

\subsection{Contributions}

In this paper, we consider the following questions:
\begin{inparaenum}[1)]
\item Under which conditions can the vector~\inputvec (and the vector \error, if appropriate) be recovered \emph{perfectly} from the (sparsely corrupted) observation $\noiseout=\dicta\inputvec + \dictb\error$, and 
\item can we formulate practical recovery algorithms with corresponding (analytical) performance guarantees? 
\end{inparaenum}
Sparsity of the signal vector \inputvec and the error vector \error will turn out to be key in answering these questions.
More specifically, based on an uncertainty relation for pairs of general dictionaries,
we establish recovery guarantees that depend on the number of nonzero entries in \inputvec and \error, and on the coherence parameters of the dictionaries \dicta and \dictb.
These recovery guarantees are obtained for the following different cases: 
\begin{inparaenum}[I)]
\item The support sets of both \inputvec and \error are known (prior to recovery), 
\item the support set of only \inputvec or only \error is known, 
\item the number of nonzero entries of only \inputvec or only \error is known, and 
\item nothing is known about \inputvec and \error.
\end{inparaenum}
We formulate efficient recovery algorithms and derive corresponding performance guarantees. 
%
%
\revision{Finally, we compare our analytical recovery thresholds to numerical results and we demonstrate the application of our algorithms and recovery guarantees to an image inpainting example.}

\subsection{Outline of the paper}

The remainder of the paper is organized as follows. In~\fref{sec:priorart}, we briefly review relevant previous results.
%
In \fref{sec:uncertainty_rel}, we derive a novel uncertainty relation that lays the foundation for the recovery guarantees reported in \fref{sec:main}.
A discussion of our results is provided in \fref{sec:discussion} and numerical results are presented in \fref{sec:simulations}. We conclude in~\fref{sec:conclusion}. 

\sloppy

\subsection{Notation}
\label{sec:notation}
Lowercase boldface letters stand for column vectors and uppercase boldface letters designate matrices. For the matrix \bM, we denote its transpose and conjugate transpose by $\bM^T$ and  $\herm{\bM}$, respectively, its (Moore--Penrose) pseudo-inverse by $\bM^{\dagger}=\left(\bM^H\bM\right)^{\!-1}\bM^H$, its $k$th column by~$\bmm_k$, and  the entry in the $k$th row and $\ell$th column by $[\bM]_{k,\ell}$. The $k$th entry of the vector $\bmm$ is $[\vecm]_k$.
The space spanned by the columns of \bM is denoted by $\setR(\bM)$.
%
%
The $M\times M$ identity matrix is denoted by $\bI_M$, the $M\times N$ all zeros matrix by $\mathbf{0}_{M,N}$, and the all-zeros vector of dimension $M$ by $\mathbf{0}_{M}$. The $M\times M$ discrete Fourier transform matrix $\bF_M$ is defined as 
\begin{align} \label{eq:fouriermatrix}
[\bF_M]_{k,\ell} = \frac{1}{\sqrt{M}} \exp \!\left(\!-\frac{2\pi i(k-1)(\ell-1)}{M}\right)\!,\,\, k,\ell\!=\!1,\ldots,M
\end{align}
where $i^2=-1$.
The Euclidean (or \elltwo) norm of the vector $\bmx$ is denoted by $\normtwo{\bmx}$, $\normone{\bmx}$ stands for the \ellone-norm of \inputvec, and $\normzero{\bmx}$ designates the number of nonzero entries in \bmx. 
Throughout the paper, we assume that the columns of the dictionaries \dicta and \dictb have unit \elltwo-norm. 
The minimum and maximum eigenvalue of the positive-semidefinite matrix \bM is denoted by \mbox{$\lambda_{\textrm{min}}(\bM)$} and $\lambda_\textrm{max}(\bM)$, respectively.
The spectral norm of the matrix \bM is $\vecnorm{\bM}=\sqrt{\lambda_\text{max}(\bM^H\bM)}$. 
Sets are designated by upper-case calligraphic letters; the cardinality of the set \setT is $\abs{\setT}$. The complement of a set~$\setS$ (in some superset~$\setT$) is denoted by $\setS^c$.
%
For two sets $\setS_1$ and $\setS_2$, $s\in\big(\setS_1+\setS_2\big)$ means that $s$ is of the form $s=s_1+s_2$, where $s_1\in\setS_1$ and $s_2\in\setS_2$.
%
%
\revision{The support set of the vector~\vecm is designated by $\supp(\vecm)$.}
%
The matrix $\bM_\setT$ is obtained from \bM by retaining the columns of \bM with indices in \setT; the vector $\bmm_\setT$ is obtained analogously. 
We define the $N\times N$ diagonal (projection) matrix~$\bP_\setS$ for the set $\setS\subseteq\{1,\ldots,N\}$ as follows:
\begin{align*}
[\bP_\setS]_{k,\ell} = 
\left\{\begin{array}{ll}
1, & k=\ell \text{ and } k \in \setS \\
0, & \textrm{otherwise.}
\end{array}\right.
\end{align*}
For $x\in\reals$,  we set $\pos{x}\!=\max\{x,0\}$. 

\fussy

\section{Review of Relevant Previous Results}
\label{sec:priorart}

Recovery of the vector \inputvec from the sparsely corrupted measurement $\noiseout = \dicta\inputvec + \dictb \error$ corresponds to a sparse-signal recovery problem subject to \emph{structured} (i.e., sparse) noise.
In this section, we briefly review relevant existing results for sparse-signal recovery from noiseless measurements, and we summarize the results available for recovery in the presence of unstructured and structured noise. 


\subsection{Recovery in the noiseless case}

Recovery of \inputvec from $\noiseout=\dicta\inputvec$ where \dicta is redundant (i.e., $\outputdim<N_a$) amounts to solving an underdetermined linear system of equations. Hence, there are infinitely many solutions \inputvec, in general.
However, under the assumption of \inputvec being sparse, the situation changes drastically. 
More specifically, one can recover \inputvec  from the observation $\noiseout=\dicta\inputvec$ by solving
\begin{align*}
(\Pzero) & \quad \trm{minimize } 
\normzero{\inputvec} \quad \trm{subject to} \quad  \noiseout=\dicta\inputvec.
\end{align*}
This approach results, however, in prohibitive computational complexity, even for small problem sizes. 
Two of the most popular and computationally tractable alternatives to solving (\Pzero) by an exhaustive search are basis pursuit (BP)~\cite{chen1998,donoho2001,donoho2002,gribonval2003,elad2002,tropp2004} and orthogonal matching pursuit (OMP)~\cite{tropp2004,Pati1993,davis1994}. 
%
%
%
BP is essentially a convex relaxation of (\Pzero) and amounts to solving
\begin{align*}
(\text{BP})  & \quad \trm{minimize } 
\normone{\inputvec} \quad \trm{subject to} \quad  \noiseout=\dicta\inputvec.
\end{align*}
OMP is a greedy algorithm that recovers the vector \inputvec by iteratively selecting the column of \dicta that is most ``correlated'' with the difference between \noiseout and its current best (in $\elltwo$-norm sense) approximation.

The questions that arise naturally are: Under which conditions does (P0) have a unique solution and when do BP and/or OMP deliver this solution? 
To formulate the answer to these questions, define $\nx=\normzero{\inputvec}$ and the coherence of the dictionary~$\dicta$ as
\begin{align} \label{eq:coherenceA}
\coha =\max_{k,\ell,k\neq \ell} \, \abs{\veca^H_k\veca_\ell}.
\end{align}
As shown in~\cite{donoho2002,gribonval2003,tropp2004}, a sufficient condition for \inputvec to be the unique solution of (\Pzero) applied to $\noiseout=\dicta\inputvec$ and for BP and OMP to deliver this solution is 
\begin{align} \label{eq:classicalthreshold}
\nx<\frac{1}{2}\Big(1+\coha^{-1}\Big).
\end{align}

\subsection{Recovery in the presence of unstructured noise}
\revision{Coherence-based recovery guarantees in the presence of unstructured (and deterministic) noise, i.e., for $\noiseout=\dicta\inputvec+\noise$, with no constraints imposed on \noise apart from $\normtwo{\noise}<\infty$, were derived in~\cite{donoho2006a,fuchs2006,tropp2006,benhaim2010,CWX10} and the references therein.}
Specifically, it was shown in~\cite{CWX10} that a suitably modified version of BP, referred to as BP denoising (BPDN), recovers an estimate $\hat{\inputvec}$ satisfying 
\mbox{$\normtwo{\bmx-\hat{\bmx}}<C\normtwo{\noise}$} \revision{provided that \fref{eq:classicalthreshold} is met.} Here, $C>0$ depends on the coherence \coha and on the sparsity level $\nx$ of \inputvec. 
\revision{Note that the support set of the estimate $\hat{\inputvec}$ may differ from that of~\inputvec.}
Another result, reported in~\cite{donoho2006a}, states that OMP delivers the correct support set (but does not perfectly recover the nonzero entries of \inputvec) provided that
\begin{align} \label{eq:noisethreshold}
\nx < \frac{1}{2}\Big(1+\coha^{-1}\Big) - \frac{\normtwo{\noise}}{\coha \abs{x_\textrm{min}}}
\end{align}
where $\abs{x_\textrm{min}}$ denotes the absolute value of the component of \inputvec with smallest nonzero magnitude.
%
The recovery condition \fref{eq:noisethreshold} yields sensible results only if $\normtwo{\noise}\!/\abs{x_\text{min}}$ is small. 
Results similar to those reported in~\cite{donoho2006a} were obtained in~\cite{fuchs2006,tropp2006}.
Recovery guarantees in the case of stochastic noise \noise can be found in~\cite{tropp2006,benhaim2010}.
We finally point out that \emph{perfect} recovery of \inputvec is, in general, impossible in the presence of \emph{unstructured} noise.
In contrast, as we shall see below, \emph{perfect} recovery is possible under structured noise according to~\fref{eq:systemmodel}.

\subsection{Recovery guarantees in the presence of structured noise}

As outlined in the introduction, many practically relevant signal recovery problems can be formulated as (sparse) signal recovery from sparsely corrupted measurements, a problem that seems to have received comparatively little attention in the literature so far and does not appear to have been developed systematically.
\begin{figure*}[!t]
\normalsize
\setcounter{MYtempeqncnt}{\value{equation}}
\setcounter{equation}{9}
\begin{equation}
\label{eq:uncertaintyconcentrated}
\abs{\setP}\abs{\setQ}\geq\frac{\pos{(1+\coha)(1-\epsilon_\setP)-\abs{\setP}\coha}\pos{(1+\cohb)(1-\epsilon_\setQ)-\abs{\setQ}\cohb}}{\cohm^2}.
\end{equation}
\setcounter{equation}{\value{MYtempeqncnt}}
\hrulefill
\vspace*{4pt}
\end{figure*}
%
%
%
%
%
%
%


\revision{
A straightforward way leading to recovery guarantees in the presence of structured noise, as in \eqref{eq:systemmodel}, follows from rewriting~\fref{eq:systemmodel} as
\begin{align} \label{eq:simpleconcatenation}
  \vecz = \dicta\inputvec+\dictb\error 
  = \dict\vecw
\end{align}
with the concatenated dictionary $\dict=\dictab$ and the stacked vector $\vecw=[\,\inputvec^T\,\,\error^T\,]^T$.
This formulation allows us to invoke the recovery guarantee in \fref{eq:classicalthreshold} for the concatenated dictionary $\dict$, which delivers a sufficient condition for~$\vecw$ (and hence, \inputvec and \error) to be the unique solution of (\Pzero) applied to $\vecz=\dict\vecw$ and for BP and OMP to deliver this solution \cite{donoho2002,gribonval2003}.
However, the so obtained recovery condition 
\begin{align} \label{eq:straightforwardABcondition}
n_w=\nx+\nerr<\frac{1}{2}\Big(1+\coh^{-1}\Big)
\end{align} 
with the  dictionary coherence \coh defined as 
\begin{align} \label{eq:coherenceAB}
\coh =\max_{k,\ell,k\neq \ell} \, \abs{\vecd^H_k\vecd_\ell}
\end{align}
\emph{ignores} the structure of the recovery problem at hand, i.e., is agnostic to i) the fact that \dict consists of the dictionaries \dicta and \dictb with known coherence parameters \coha and \cohb, respectively, and ii)  knowledge about the support sets of \inputvec and/or \error that may be available prior to recovery.
As shown in \fref{sec:main}, exploiting these two structural aspects of the recovery problem yields superior (i.e., less restrictive) recovery thresholds.}
\revision{Note that condition~\fref{eq:straightforwardABcondition} guarantees perfect recovery of \inputvec (and \error) independent of the \elltwo-norm of the noise vector, i.e., $\normtwo{\dictb\error}$ may be arbitrarily large.  
This is in stark contrast to the recovery guarantees for noisy measurements in~\cite{CWX10} and \fref{eq:noisethreshold} (originally reported in~\cite{donoho2006a}).}

\revision{Special cases of the general setup~\fref{eq:systemmodel}, explicitly taking into account certain structural aspects of the recovery problem were considered in~\cite{donoho1989,laska2009democracy,laska,candes2005decoding,carrillo2010,wright2010,nguyen2011}.
%
Specifically, in~\cite{donoho1989} it was shown that for $\dicta=\bF_\outputdim$, $\dictb=\bI_\outputdim$, and knowledge of the support set of \error, \emph{perfect} recovery of the $\outputdim$-dimensional vector \inputvec is possible if 
\begin{align} \label{eq:donohostarkcondition}
  2\nx n_e < \outputdim
\end{align}
where $n_e=\normzero{\error}$.}
%
%
%
%
\revision{In~\cite{laska2009democracy,laska}, recovery guarantees based on the RIC of the matrix \dicta for the case where \dictb is an orthonormal basis (ONB), and where the support set of \error is either known or unknown, were reported; these recovery guarantees are particularly handy when \dicta is, for example, i.i.d.~Gaussian~\cite{donoho2006,candes2006}.} 
\revision{However, results for the case of \dicta and \dictb both \emph{general} (and deterministic) dictionaries taking into account prior knowledge about the support sets of \inputvec and \error seem to be missing in the literature.}
%
%
\revision{Recovery guarantees for~\dicta i.i.d.~non-zero mean Gaussian, $\dictb=\bI_\outputdim$, and the support sets of \vecx and \error unknown were reported in~\cite{wright2010}.}
\revision{In~\cite{nguyen2011} recovery guarantees under a probabilistic model on both \vecx and \vece and for unitary \dicta and $\dictb=\bI_{M}$ were reported showing that \inputvec can be recovered perfectly with high probability (and independently of the \elltwo-norm of \vecx and \error).}
%
%
The problem of sparse-signal recovery in the presence of impulse noise (i.e., $\dictb=\bI_\outputdim$) was considered in~\cite{carrillo2010}, where a particular nonlinear measurement process combined with a non-convex program for signal recovery was proposed. 
In \cite{candes2005decoding}, signal recovery in the presence of impulse noise  based on \ellone-norm minimization was investigated. The setup in  \cite{candes2005decoding}, however, differs considerably from the one considered in this paper as \dicta in \cite{candes2005decoding} needs to be tall (i.e., $\outputdim>N_a$) and the vector \inputvec to be recovered is not necessarily sparse.

We conclude this literature overview by noting that the present paper is 
inspired by~\cite{donoho1989}. 
Specifically, we note that the recovery guarantee \fref{eq:donohostarkcondition} reported  in~\cite{donoho1989} is obtained from an uncertainty relation that puts limits on how sparse a given signal can simultaneously be in the Fourier basis and in the identity basis.
Inspired by this observation, we start our discussion by presenting an uncertainty relation for pairs of general dictionaries, which forms the basis for the recovery guarantees reported later in this paper. 
 
\section{A General Uncertainty Relation for $\epsilon$-Concentrated Vectors}\label{sec:uncertainty_rel}

We next present a novel uncertainty relation, which extends the uncertainty relation in~\cite[Lem.~1]{kuppinger2010a} for pairs of general dictionaries to vectors that are $\epsilon$-\emph{concentrated} rather than perfectly sparse. 
\revision{As shown in \fref{sec:main}, this extension constitutes the basis for the derivation of recovery guarantees for BP.}

\subsection{The uncertainty relation}

Define the mutual coherence between the dictionaries \dicta and \dictb as
\begin{align*}
\cohm = \max_{k,\ell} \, \abs{\veca^H_k\vecb_\ell}.
\end{align*}
Furthermore, we will need the following definition, which appeared previously in~\cite{donoho1989}.
\begin{defi}
A vector $\vecr\in\complexset^{\inputdim_r}$ is said to be $\epsilon_{\setR}$-\emph{concentrated} to the set $\setR\subseteq \{1,\ldots,\inputdim_r\}$ if $\normone{\bP_\setR\vecr}\geq(1-\epsilon_\setR)\normone{\vecr}$, where $\epsilon_\setR\in[0,1]$. We say that the vector $\vecr$ is \emph{perfectly concentrated} to the set $\setR$ and, hence, $\abs{\setR}$-sparse if $\bP_\setR\vecr = \vecr$, i.e., if $\epsilon_{\setR} = 0$.
\end{defi}

%
%
We can now state the following uncertainty relation for pairs of general dictionaries and for $\epsilon$-concentrated vectors.
\begin{thm}\label{thm:uncertainty}
Let $\dicta\in\complexset^{\outputdim\times\inputdimA}$ be a dictionary with coherence $\coha$, $\dictb\in\complexset^{\outputdim\times\inputdimB}$ a dictionary with coherence \cohb, and denote the mutual coherence between \dicta and \dictb by \cohm.
Let $\vecs$ be a vector in~$\complexset^\outputdim$ that can be represented as a linear combination of columns of \dicta and, \revision{similarly}, as a linear combination of columns of \dictb. 
Concretely, there exists a pair of vectors $\vecp\in\complexset^\inputdimA$ and $\vecq\in\complexset^\inputdimB$ such that $\vecs=\dicta\vecp=\dictb\vecq$ (we exclude the trivial case where $\vecp=\bZero_{\inputdimA}$ and $\vecq=\bZero_{\inputdimB}$).\footnote{The uncertainty relation continues to hold if either $\vecp=\bZero_{N_a}$ or $\vecq=\bZero_{N_b}$, but does not apply to the trivial case $\vecp=\bZero_{N_a}$ \emph{and} $\vecq=\bZero_{N_b}$. In all three cases we have $\vecs=\bZero_{\outputdim}$.}
%
%
%
%
%
%
%
%
If $\vecp$ is $\epsilon_\setP$-concentrated to \setP and $\vecq$ is $\epsilon_\setQ$-concentrated to $\setQ$,  then \fref{eq:uncertaintyconcentrated} holds. \addtocounter{equation}{1}
\end{thm}
\begin{IEEEproof}
\revision{The proof follows closely that of \cite[Lem.~1]{kuppinger2010a}, which applies to perfectly concentrated vectors \vecp and \vecq.
We therefore only summarize the modifications to the proof of \cite[Lem.~1]{kuppinger2010a}.
Instead of using $\sum_{p\in\setP}\abs{[\vecp]_p}=\normone{\vecp}$ to arrive at \cite[Eq.~29]{kuppinger2010a}
\begin{align*}
\pos{(1+\coha)-\abs{\setP}\coha}\normone{\vecp}\leq\abs{\setP}\cohm\normone{\vecq}
\end{align*}
we invoke $\sum_{p\in\setP}\abs{[\vecp]_p}\geq(1-\epsilon_\setP)\normone{\vecp}$ to arrive at the following inequality valid for $\epsilon_\setP$-concentrated vectors \vecp:
\begin{align}\label{eq:uncertain5}
\pos{(1+\coha)(1-\epsilon_\setP)-\abs{\setP}\coha}\normone{\vecp}\leq\abs{\setP}\cohm\normone{\vecq}.
\end{align}
Similarly, $\epsilon_\setQ$-concentration, i.e., $\sum_\setQ\abs{[\vecq]_q}\geq(1-\epsilon_\setQ)\normone{\vecq}$, is used to replace \cite[Eq.~30]{kuppinger2010a} by 
\begin{align} \label{eq:uncertain6}
 \pos{(1+\cohb)(1-\epsilon_\setQ)-\abs{\setQ}\cohb}\normone{\vecq}\leq\abs{\setQ}\cohm\normone{\vecp}.
\end{align}
The uncertainty relation \fref{eq:uncertaintyconcentrated} is then obtained by multiplying \fref{eq:uncertain5} and \fref{eq:uncertain6} and dividing the resulting inequality by $\normone{\vecp}\normone{\vecq}$.}
\end{IEEEproof}

\sloppy

%
%
%
%
\revision{In the case where both \vecp and \vecq are \emph{perfectly concentrated}, i.e., $\epsilon_\setP=\epsilon_\setQ=0$, \fref{thm:uncertainty} reduces to the uncertainty relation reported in~\cite[Lem.~1]{kuppinger2010a}, which we restate next for the sake of completeness.} 
\begin{cor}[{\!\!\cite[Lem.~1]{kuppinger2010a}}]\label{cor:uncertainty}
If $\setP=\supp(\vecp)$ and $\setQ=\supp(\vecq)$,
the following holds:
%
\begin{align}\label{eq:uncertainty}
\abs{\setP}\abs{\setQ}\geq \frac{\pos{1-\coha\!\left(\abs{\setP}-1\right)}\pos{1-\cohb\!\left(\abs{\setQ}-1\right)}}{\cohm^2}. 
\end{align}
\end{cor}
As detailed in~\cite{kuppinger2010a,Kuppinger2011}, the uncertainty relation in \fref{cor:uncertainty} generalizes the uncertainty relation for two orthonormal bases (ONBs) found in~\cite{elad2002}. Furthermore, it extends the uncertainty relations provided in~\cite{ghobber2010} for pairs of square dictionaries (having the same number of rows and columns) to pairs of general dictionaries \dicta and~\dictb. 

\fussy

\subsection{Tightness of the uncertainty relation}
\label{sec:uncertainty_tightness}
\revision{In certain special cases it is possible to find signals that satisfy the uncertainty relation~\fref{eq:uncertaintyconcentrated} with equality.}
As in \cite{donoho1989}, consider $\dicta=\bF_\outputdim$ and $\dictb=\bI_\outputdim$, so that $\cohm=1/\sqrt{\outputdim}$, and define the comb signal containing equidistant spikes of unit height as 
\be
	[\comb{t}]_\indexa=\left\{\begin{array}{ll} 1, & \trm{if } (\indexa-1)\trm{ mod } t=0\\
	0,& \trm{otherwise}\end{array}\right.
\een
where we shall assume that $t$ divides \outputdim.
It can be shown that the vectors $\vecp=\comb{\sqrt{\outputdim}}$ and $\vecq=\comb{\sqrt{\outputdim}}$, both having $\sqrt{\outputdim}$ nonzero entries, satisfy $\bF_\outputdim\vecp=\bI_\outputdim \vecq$. 
If $\setP = \supp(\vecp)$ and  $\setQ = \supp(\vecq)$, the vectors \vecp and \vecq are perfectly concentrated to \setP and \setQ, respectively, i.e., $\epsilon_\setP=\epsilon_\setQ=0$. 
Since $\abs{\setP}=\abs{\setQ} = \sqrt{\outputdim}$ and $\cohm=1/\sqrt{\outputdim}$ it follows that  $\abs{\setP}\abs{\setQ} = 1/\cohm^2 = M$ and, \revision{hence, $\vecp=\vecq=\comb{\sqrt{\outputdim}}$ satisfies~\fref{eq:uncertaintyconcentrated} with equality.}

\revision{We will next show that for pairs of general dictionaries \dicta and \dictb, finding signals that satisfy the uncertainty relation \fref{eq:uncertaintyconcentrated} with equality is NP-hard.}
For the sake of simplicity, we restrict ourselves to the case $\setP = \supp(\vecp)$ and  $\setQ = \supp(\vecq)$, which implies $\abs{\setP} = \normzero{\vecp}$ and $\abs{\setQ} = \normzero{\vecq}$. 
Next, consider the problem
\begin{align*} 
 (\textrm{U}0) \quad 
\left\{ \begin{array}{ll}
 \text{minimize} & 
 \normzero{\vecp}\normzero{\vecq}  \\[0.1cm] 
 \trm{subject to} &  \dicta\vecp=\dictb\vecq, \,\, \normzero{\vecp}\ge 1, \normzero{\vecq}\ge 1.
 \end{array} \right.
\end{align*}
Since we are interested in the minimum of  $\normzero{\vecp}\normzero{\vecq}$ for nonzero vectors \vecp and \vecq, we imposed the constraints $\normzero{\vecp}\ge 1$ and $ \normzero{\vecq}\ge 1$ to exclude the case where $\vecp=\bZero_{N_a}$ and/or $\vecq=\bZero_{N_b}$.
Now, it follows that for the particular choice $\dictb=\noiseout\in\complexset^\outputdim$ and hence $\vecq=q\in\complexset\setminus\{0\}$ (note that we exclude the case $q=0$ as a consequence of the requirement $\normzero{\vecq}\geq1$) the problem $(\textrm{U}0)$ reduces to
\begin{align*}
(\textrm{U}0^*) & \quad \trm{minimize } 
\normzero{\inputvec} \quad \trm{subject to} \quad  \dicta\inputvec=\noiseout
\end{align*}
where $\inputvec=\vecp/q$. \revision{However, as $(\textrm{U}0^*)$ is equivalent to (\Pzero), which is NP-hard~\cite{davis1997}, in general, we can conclude that finding a pair $\vecp$ and $\vecq$ satisfying the uncertainty relation~\eqref{eq:uncertaintyconcentrated} with equality is NP-hard.}
%


\section{Recovery of Sparsely Corrupted Signals}\label{sec:main}
Based on the uncertainty relation in \fref{thm:uncertainty}, we next derive conditions that guarantee perfect recovery of  \inputvec (and of \error, if appropriate) from the (sparsely corrupted) measurement $\vecz=\dicta\inputvec+\dictb\error$. 
These conditions will be seen to depend on the number of nonzero entries of \inputvec and \error, and on the coherence parameters \coha, \cohb, and \cohm.
Moreover, in contrast to~\fref{eq:noisethreshold}, the recovery conditions we find will not depend on the \elltwo-norm of the noise vector~$\normtwo{\dictb\error}$, which is hence allowed to be arbitrarily large.
We consider the following cases:
\begin{inparaenum}[I)]
\item The support sets of both \inputvec and \error are known (prior to recovery), 
\item the support set of only \inputvec or only \error is known, 
\item the number of nonzero entries of only \inputvec or only \error is known, and 
\item nothing is known about \inputvec and \error.
\end{inparaenum}
The uncertainty relation in~\fref{thm:uncertainty} is the basis for the recovery guarantees in all four cases considered.
%
%
To simplify notation, motivated by the form of the right-hand side (RHS) of \fref{eq:uncertainty}, we define the function
\begin{align*}
  f(u,v) = \frac{\pos{1-\coha\!\left(u-1\right)}\pos{1-\cohb\!\left(v-1\right)}}{\cohm^2}.
\end{align*}
In the remainder of the paper, \setX denotes $\supp(\inputvec)$ and \setE stands for $\supp(\error)$.
We furthermore assume that the dictionaries $\dicta$ and $\dictb$ are known perfectly to the recovery algorithms. 
Moreover, we assume that\footnote{If $\cohm=0$, the space spanned by the columns of \dicta is orthogonal to the space spanned by the columns of \dictb. 
This makes the separation of the components $\dicta\inputvec$ and $\dictb\error$ given \noiseout straightforward.  Once this separation is accomplished, \inputvec can be recovered from $\dicta\inputvec$ using (P0), BP, or OMP, if \fref{eq:classicalthreshold} is satisfied.} $\cohm>0$.

\subsection{Case I: Knowledge of \setX and \setE}
\label{sec:bothknown}

We start with the case where both \setX and \setE are known prior to recovery. The values of the nonzero entries of \inputvec and \error are unknown. 
This scenario is relevant, for example, in applications requiring recovery of clipped band-limited signals with known spectral support $\setX$. Here, we would have $\dicta=\bF_\outputdim$, $\dictb=\bI_\outputdim$, and \setE can be determined as follows:  Compare the measurements~$[\noiseout]_i$, $i=1,\ldots,\outputdim$, to the clipping threshold~$a$; if $\abs{[\noiseout]_i} = a$ add the corresponding index $i$ to \setE. 

Recovery of \inputvec from \noiseout is then performed as follows. 
We first rewrite the input-output relation in~\fref{eq:systemmodel} as
\begin{align*}
\noiseout = \dicta_\setX \inputvec_\setX + \dictb_\setE \error_\setE = \dict_{\setX,\setE} \bms_{\setX,\setE}
\end{align*}
with the concatenated dictionary $\dict_{\setX,\setE} = \left[\,\bA_\setX\,\,\bB_\setE\,\right]$ and the stacked vector $\bms_{\setX,\setE} = \big[\,\inputvec_\setX^T\,\,\error_\setE^T\,\big]^T$. 
%
Since \setX and \setE are known, we can recover the stacked vector $\bms_{\setX,\setE} = \big[\,\inputvec_\setX^T\,\,\error_\setE^T\,\big]^T$, perfectly and, hence, the nonzero entries of both \inputvec and \error, if the pseudo-inverse $\dict^\dagger_{\setX,\setE}$ exists. In this case, we can obtain $\bms_{\setX,\setE}$, as
\begin{align} \label{eq:recoveryifeverythingisknown}
\bms_{\setX,\setE} = \dict^\dagger_{\setX,\setE} \noiseout.
\end{align}
The following theorem states a sufficient condition for  $\dict_{\setX,\setE}$ to have full (column) rank, which  implies existence of the pseudo-inverse $\dict^\dagger_{\setX,\setE}$.
This condition depends on the coherence parameters \coha, \cohb, and \cohm, of the involved dictionaries \dicta and \dictb and on \setX and \setE  through the cardinalities $\abs{\setX}$ and $\abs{\setE}$, i.e., the \emph{number of nonzero entries} in \inputvec and \error, respectively.
\begin{thm}\label{thm:P1_bothknown}
Let $\noiseout=\dicta\inputvec+\dictb\error$ with $\setX=\supp(\inputvec)$ and $\setE=\supp(\error)$. Define $\nx=\normzero{\inputvec}$ and $n_e=\normzero{\error}$. If 
\begin{align}\label{eq:P1_bothknown_assump}
  \nx n_e<f(\nx,n_e),
\end{align}
then the concatenated dictionary $\dict_{\setX,\setE} = \left[\,\bA_\setX\,\,\bB_\setE\,\right]$ has full (column) rank. 
\end{thm}
\begin{IEEEproof}
 See \fref{app:P1_bothknown_proof}. 
 \end{IEEEproof} 
 %
%

For the special case $\dicta=\bF_\outputdim$ and $\dictb=\bI_\outputdim$ (so that $\coha=\cohb=0$ and $\cohm=1/\sqrt{\outputdim}$) the recovery condition \fref{eq:P1_bothknown_assump} reduces to $\nx n_e<\outputdim$, a result obtained previously in~\cite{donoho1989}. 
\revision{Tightness of \fref{eq:P1_bothknown_assump} can be established by noting that the pairs $\inputvec = \lambda \comb{\sqrt{\outputdim}}$, $\bme = (1-\lambda)\comb{\sqrt{\outputdim}}$ with $\lambda\in(0,1)$ and $\inputvec' = \lambda' \comb{\sqrt{\outputdim}}$, $\bme' = (1-\lambda')\comb{\sqrt{\outputdim}}$ with $\lambda'\neq\lambda$ and  $\lambda'\in(0,1)$ both satisfy \fref{eq:P1_bothknown_assump} with equality and lead to the same measurement outcome $\vecz=\bF_M\vecx+\vece=\bF_M\vecx'+\vece'$ \cite{Kuppinger2011}.}


\revision{It is interesting to observe that \fref{thm:P1_bothknown} yields a sufficient condition on \nx and $n_e$ for any $(\outputdim-n_e)\times\nx$-submatrix of \dicta to have full (column) rank. 
To see this, consider the special case $\dictb=\bI_\outputdim$
and hence, $\dict_{\setX,\setE}=\left[\,\bA_\setX\,\,\bI_\setE\,\right]$.}
Condition \fref{eq:P1_bothknown_assump} characterizes pairs ($\nx,n_e$), for which all matrices $\dict_{\setX,\setE}$ with $\nx=\abs{\setX}$ and $n_e=\abs{\setE}$ are guaranteed to have full (column) rank.
Hence, the sub-matrix consisting of all rows of $\dicta_\setX$ with row index in $\setE^c$ must have full (column) rank as well. Since the result holds for all support sets \setX and \setE with $\abs{\setX}=\nx$ and $\abs{\setE}=n_e$, all possible $(\outputdim-n_e)\times\nx$-submatrices of~\dicta must have full (column) rank.

\subsection{Case II: Only \setX or only \setE is known}
\label{sec:supportKnown}
Next, we find recovery guarantees for the case where either only \setX or only \setE is known prior to recovery.

\subsubsection{Recovery when \setE is known and \setX is unknown}
\label{sec:Eknown}
A prominent application for this setup is the recovery of clipped band-limited signals~\cite{Adler2011,laska2009democracy}, where the signal's spectral support, i.e., \setX, is unknown. The support set \setE 
can be identified as detailed previously in \fref{sec:bothknown}.
Further application examples for this setup include inpainting and super-resolution~\cite{mallat2010,elad2001,bertalmio2000} of signals that admit a sparse representation in \dicta (but with unknown support set \setX). 
%
%
The locations of the missing elements in $\outputvec=\dicta\inputvec$ are known (and correspond, e.g., to missing paint elements in frescos), i.e., the set \setE can be determined prior to recovery.
Inpainting and super-resolution then amount to reconstructing the vector \inputvec from the sparsely corrupted measurement $\noiseout = \dicta\inputvec + \error$ and computing $\outputvec=\dicta\inputvec$.


%

\sloppy

The setting of \setE known and \setX unknown was considered previously in~\cite{donoho1989} for the special case $\dicta=\bF_\outputdim$ and $\dictb=\bI_\outputdim$. The recovery condition \fref{eq:P0uk_assump} in \fref{thm:P0_uk} below extends the result in~\cite[Thms.~5 and~9]{donoho1989} to pairs of general dictionaries \dicta and~\dictb. 
\begin{thm}\label{thm:P0_uk}
Let $\noiseout=\dicta\inputvec+\dictb\error$ where $\setE=\supp(\error)$ is known. Consider the problem
\begin{align} \label{eq:P0uk_problem}
 (\Pzero,\setE) \quad \left\{
 \begin{array}{ll}
 \text{minimize} &
 \normzero{\tilde{\inputvec}} \\[0.1cm]
  \trm{subject to} & \dicta\tilde{\inputvec} \in \left(\left\{ \noiseout\right\} + \setR(\dictb_\setE)\right)
\end{array}\right.
\end{align}
and the convex program 
\begin{align} \label{eq:P1uk_problem}
 (\textrm{BP},\setE) \quad \left\{
 \begin{array}{ll}
 \text{minimize} &
 \normone{\tilde{\inputvec}} \\[0.1cm]
  \trm{subject to} & \dicta\tilde{\inputvec} \in \left(\left\{ \noiseout\right\} + \setR(\dictb_\setE)\right) .
 \end{array} \right.
\end{align}
If 
$\nx=\normzero{\inputvec}$ and $n_e=\normzero{\error}$ satisfy
\begin{align}\label{eq:P0uk_assump}
 2\nx n_e<f(2\nx,n_e),
\end{align}
then the unique solution of $(\Pzero,\setE)$ applied to $\noiseout=\dicta\inputvec+\dictb\error$ is given by \inputvec and $(\textrm{BP},\setE)$ will deliver this solution.
\end{thm}
\begin{IEEEproof}
See~\fref{app:P0P1_uk_proof}.
\end{IEEEproof}
Solving $(\Pzero,\setE)$ requires a combinatorial search, which results in prohibitive computational complexity even for moderate problem sizes. The convex relaxation $(\textrm{BP},\setE)$ can, however, be solved more efficiently.
Note that the constraint $ \dicta\tilde{\inputvec} \in \left( \left\{ \noiseout\right\} + \setR(\dictb_\setE) \right)$ reflects the fact that any error component $\dictb_\setE\error_\setE$ yields consistency on account of \setE known (by assumption).
%
%
\revision{For $n_e=0$ (i.e., the noiseless case) the recovery threshold~\eqref{eq:P0uk_assump} reduces to $\nx<(1+1/\coha)/2$, which is the well-known recovery threshold \fref{eq:classicalthreshold} guaranteeing recovery of the sparse vector \inputvec through (\Pzero) and BP applied to $\noiseout=\dicta\inputvec$.}
\revision{We finally note that RIC-based guarantees for recovering \inputvec from  $\noiseout=\dicta\inputvec$ (i.e., recovery in the absence of (sparse) corruptions) that take into account partial knowledge of the signal support set $\setX$ were developed in \cite{VL2010,Jacques2010}.}

\revision{Tightness of \fref{eq:P0uk_assump} can be established by setting $\dicta=\bF_\outputdim$ and $\dictb=\bI_\outputdim$.  Specifically, the pairs $\inputvec=\comb{2\sqrt{\outputdim}}-\comb{\sqrt{\outputdim}}$, $\error=\comb{\sqrt{\outputdim}}$  and $\inputvec'=\comb{2\sqrt{\outputdim}}$, $\vece'=\vece$ both satisfy \fref{eq:P0uk_assump} with equality.
One can furthermore verify that $\inputvec$ and $\inputvec'$ are both in the admissible set specified by the constraints in $(\Pzero,\setE)$ and $(\textrm{BP},\setE)$ and $\normzero{\inputvec'}=\normzero{\inputvec}$, $\normone{\inputvec'}=\normone{\inputvec}$. Hence, $(\Pzero,\setE)$ and $(\textrm{BP},\setE)$ both cannot distinguish between $\vecx$ and $\vecx'$ based on the measurement outcome \vecz. For a detailed discussion of this example we refer to \cite{Kuppinger2011}.}

\fussy

Rather than solving $(\Pzero,\setE)$ or $(\text{BP},\setE)$, we may attempt to recover the vector \inputvec by exploiting more directly the fact that $\setR(\dictb_\setE)$ is known (since \dictb and \setE are assumed to be known) and projecting the measurement outcome \noiseout onto the orthogonal complement of $\setR(\dictb_\setE)$. 
%
This approach would eliminate the (sparse) noise component and leave us with a standard 
sparse-signal recovery problem for the vector \inputvec.
We next show that this ansatz is guaranteed to recover the sparse vector \inputvec provided that  condition~\eqref{eq:P0uk_assump} is satisfied.
%
%
%
Let us detail the procedure.
If the columns of $\dictb_\setE$ are linearly independent, the pseudo-inverse $\dictb^\dagger_\setE$ exists, and the projector onto the orthogonal complement of $\setR(\dictb_\setE)$ is given by
%
\begin{align} \label{eq:deletionmatrix}
\bR_\setE = \bI_{\outputdim} - \dictb_\setE \dictb^\dagger_\setE.
\end{align}
Applying $\bR_\setE$ to the measurement outcome \noiseout yields
\begin{align} \label{eq:modifiedconstraint}
 \bR_\setE \vecz &= \bR_\setE \!\left(\dicta \inputvec+\dictb_\setE\error_\setE\right) = \bR_\setE \dicta \inputvec\triangleq\hat{\noiseout} 
\end{align}
where we used the fact that $\bR_\setE\dictb_\setE=\mathbf{0}_{\outputdim,n_e}$.
We are now left with the standard problem of recovering \inputvec from the modified measurement outcome $\hat{\vecz}=\bR_\setE \dicta \inputvec$.
What comes to mind first is that computing the standard recovery threshold \fref{eq:classicalthreshold} for the modified dictionary $\bR_\setE \dicta$ should provide us with a recovery threshold for the problem of extracting \inputvec from $\hat{\vecz}=\bR_\setE \dicta \inputvec$.
%
%
It turns out, however, that the columns of $\bR_\setE \dicta$ will, in general, not have unit \elltwo-norm, an assumption underlying~\eqref{eq:classicalthreshold}. 
What comes to our rescue is that under condition~\eqref{eq:P0uk_assump} we have (as shown in~\fref{thm:P1uk_equivalence} below) $\normtwo{\bR_\setE \cola_\ell}>0$ for $\allonotwo{\ell}{\inputdimA}$. 
%
We can, therefore, normalize the modified dictionary $\bR_\setE \dicta$ by rewriting~\eqref{eq:modifiedconstraint} as  
\begin{align} \label{eq:projectednormalizeddict}
	\hat{\vecz} = \bR_\setE \dicta\bDelta\hat\inputvec
\end{align}
where $\bDelta$ is the diagonal matrix with elements
%
\begin{align*}
 [\bDelta]_{\ell,\ell} = \frac{1}{\normtwo{\bR_\setE \cola_\ell}}, \quad \ell=1,\ldots,N_a,
\end{align*}
and $\hat\inputvec\triangleq\bDelta^{\!-1} \inputvec$. 
Now, $\bR_\setE \dicta\bDelta$ plays the role of the dictionary (with normalized columns) and $\hat\inputvec$ is the  unknown sparse vector that we wish to recover. 
Obviously, $\supp(\hat\inputvec)=\supp(\inputvec)$ and \inputvec can be recovered from $\hat\inputvec$ according to\footnote{If $\normtwo{\bR_\setE \cola_\ell}>0$ for $\allonotwo{\ell}{\inputdimA}$, then the matrix \bDelta corresponds to a one-to-one mapping.} $\inputvec=\bDelta \hat\inputvec$.
%
%
%
%
%
%
The following theorem shows that \fref{eq:P0uk_assump} is sufficient to guarantee the following: 
\begin{inparaenum}[i)]
\item The columns of $\dictb_\setE$ are linearly independent, which guarantees the existence of $\dictb^\dagger_\setE$, 
\item $\normtwo{\bR_\setE \cola_\ell}>0$ for $\allonotwo{\ell}{\inputdimA}$, and
\item no vector $\inputvec'\in\complexset^{\inputdimA}$ with $\normzero{\inputvec'}\leq 2\nx$ lies in the kernel of $\bR_\setE\dicta$.
\end{inparaenum}
Hence, \fref{eq:P0uk_assump} enables perfect recovery of \inputvec from \fref{eq:projectednormalizeddict}.
\begin{thm} \label{thm:P1uk_equivalence}
If \fref{eq:P0uk_assump} is satisfied, the unique solution of (\Pzero) applied to $\hat{\vecz}=\bR_\setE \dicta\bDelta \hat\inputvec$ is given by $\hat\inputvec$. Furthermore, BP and OMP applied to $\hat{\vecz}=\bR_\setE \dicta\bDelta \hat\inputvec$ are guaranteed to recover the unique (\Pzero)-solution.
\end{thm}
\begin{IEEEproof}
See \fref{app:P1_uk_equivalence_proof}.
\end{IEEEproof} 

Since condition~\fref{eq:P0uk_assump} ensures that $[\bDelta]_{\ell,\ell}>0$, $\ell=1,\ldots,N_a$, the vector \inputvec can be obtained from $\hat\inputvec$ according to $\inputvec=\bDelta\hat\inputvec$.
Furthermore, \fref{eq:P0uk_assump} guarantees the existence of $\dictb^\dagger_\setE$ and hence the nonzero entries of \error can be obtained from $\inputvec$ as follows:
\begin{align*}
\error_\setE = \dictb^\dagger_\setE\!\left(\vecz-\dicta\inputvec\right).
\end{align*}
\fref{thm:P1uk_equivalence} generalizes the results in~\cite[Thms.~5 and~9]{donoho1989} obtained for the special case $\dicta=\bF_\outputdim$ and $\dictb=\bI_\outputdim$ to  pairs of general dictionaries and additionally shows that OMP delivers the correct solution provided that \fref{eq:P0uk_assump} is satisfied.

\sloppy

%
\revision{It follows from \fref{eq:projectednormalizeddict} that other sparse-signal recovery algorithms, such as iterative thresholding-based algorithms~\cite{maleki2010}, CoSaMP~\cite{Tropp08}, or subspace pursuit~\cite{DM09} can be applied to recover~\inputvec.\footnote{\revision{Finding analytical recovery guarantees for these algorithms remains an interesting open  problem.}}}
%
Finally, we note that the idea of projecting the measurement outcome onto the orthogonal complement of the space spanned by the active columns of \dictb \revision{and investigating the effect on  the RICs, instead of the coherence parameter \coha (as was done in~\fref{app:effectivecoherence})
was put forward in~\cite{laska2009democracy,DBB09} along with RIC-based recovery guarantees that apply to random matrices~\dicta and guarantee the recovery of \inputvec with high probability (with respect to \dicta and irrespective of the locations of the sparse corruptions).}

\fussy

%

\subsubsection{Recovery when \setX is known and \setE is unknown}
A possible application scenario for this situation is the recovery of spectrally sparse signals with known spectral support that are impaired by impulse noise with unknown impulse locations.

It is evident that this setup is formally equivalent to that discussed in~\fref{sec:Eknown}, with the roles of \inputvec and \error interchanged. 
In particular, we may apply the projection matrix $\bR_\setX=\bI_\outputdim-\dicta_\setX\dicta_\setX^\dagger$ to the corrupted measurement outcome \vecz to obtain the standard recovery problem $\hat{\vecz}'=\bR_\setX \dictb\bDelta'\hat{\vece}$, where $\bDelta'$ is a diagonal matrix with diagonal elements $[\bDelta']_{\ell,\ell} = 1/\normtwo{\bR_\setX \colb_\ell}$. The corresponding unknown vector is given by $\hat\vece\triangleq(\bDelta')^{-1}\,\vece$. 
The following corollary is a direct consequence of~\fref{thm:P1uk_equivalence}.
\begin{cor}\label{cor:P1_signal_known}
Let $\noiseout=\dicta\inputvec+\dictb\error$ where $\setX=\supp(\inputvec)$ is known.
If the number of nonzero entries in \inputvec and \error, i.e., $\nx=\normzero{\inputvec}$ and $n_e=\normzero{\error}$, satisfy 
\begin{align}\label{eq:P1signaluk_assump}
  2\nx n_e<f(\nx,2n_e)
\end{align}
then the unique solution of (\Pzero) applied to $\hat{\vecz}'=\bR_\setX \dictb\bDelta' \hat\vece$ is given by $\hat\vece=(\bDelta')^{-1}\,\vece$. 
%
%
%
%
%
%
Furthermore, BP and OMP applied to $\hat{\vecz}'=\bR_\setX \dictb\bDelta' \hat\vece$ recover the unique (\Pzero)-solution.
\end{cor}

Once we have $\hat\error$, the vector \vece can be obtained easily, since $\vece=\bDelta'\hat\vece$ and the nonzero entries of \inputvec are given by 
\begin{align*}
\inputvec_\setX = \dicta^\dagger_\setX(\vecz-\dictb\error).
\end{align*}
\revision{Since~\eqref{eq:P1signaluk_assump} ensures that the columns of $\dicta_\setX$ are linearly independent, the pseudo-inverse  $\dicta^\dagger_\setX$ is guaranteed to exist.
Note that tightness of the recovery condition \fref{eq:P1signaluk_assump} can be established analogously to the case of \setE  known and \setX  unknown (discussed in~\fref{sec:Eknown}).}

\subsection{Case III: Cardinality of \setE or \setX known}
\label{sec:cardinalityKnown}
We next consider the case where neither \setX nor \setE are known, but knowledge of either $\normzero{\inputvec}$ or $\normzero{\error}$ is available (prior to recovery). 
\revision{An application scenario for $\normzero{\inputvec}$ unknown and $\normzero{\error}$ known would be the recovery of a sparse pulse-stream with unknown pulse-locations from measurements that are corrupted by electric hum with unknown base-frequency but known number of harmonics (e.g., determined by the base frequency of the hum and the acquisition bandwidth of the system under consideration).}
We state our main result for the case $n_e=\normzero{\error}$ known and $\nx=\normzero{\inputvec}$ unknown. The case where \nx is known and $n_e$ is unknown can be treated similarly.
\begin{thm}\label{thm:P0_nothingknown}
Let $\noiseout=\dicta\inputvec+\dictb\error$, define $\nx=\normzero{\inputvec}$ and $n_e=\normzero{\error}$, and assume that $n_e$ is known. Consider the problem
\begin{align} \label{eq:P0_nothinknown_problem}
(\Pzero, n_e) \quad
\lefto\{\begin{array}{ll}
\text{minimize} &
\normzero{\tilde{\inputvec}} \\
  \trm{subject to} & \dicta\tilde{\inputvec}\in \Big(\{\vecz\} +\!\!\!\bigcup\limits_{\setE'\in\mathscr{P}}\!\!\setR(\dictb_{\setE'})\Big)
\end{array}\right.
\end{align}
where $\mathscr{P} =\wp_{n_e}(\{1,\ldots,\inputdimB\})$ \revision{denotes the set of subsets of $\{1,\ldots,\inputdimB\}$ of cardinality less than or equal to $n_e$.}
The unique solution of  ($\Pzero,n_e$) applied to $\noiseout=\dicta\inputvec+\dictb\error$ is given by \inputvec if
\begin{align}\label{eq:P0_nothingknown_assump}
  4\nx n_e<f(2\nx,2n_e).
\end{align}
\end{thm}

\begin{IEEEproof}
See~\fref{app:P1_bothuk_proof}.
\end{IEEEproof}

\revision{We emphasize that the problem $(\Pzero,n_e)$ exhibits prohibitive (concretely, combinatorial) computational complexity, in general.}
Unfortunately, replacing the \ellzero-norm of $\tilde\inputvec$ in the minimization in~\eqref{eq:P0_nothinknown_problem} by the \ellone-norm does not lead to a computationally tractable alternative either, as the constraint $ \dicta\tilde{\inputvec}\in (\{\vecz\} +\bigcup_{\setE'\in\mathscr{P}}\setR(\dictb_{\setE'}))$ specifies a non-convex set, in general. 
Nevertheless, the recovery threshold in~\eqref{eq:P0_nothingknown_assump} is interesting as it completes the picture on the impact of knowledge about the support sets of \inputvec and \error on the recovery thresholds. We refer to \fref{sec:factortwoinguarantees} for a detailed discussion of this matter. 
\revision{Note, though, that greedy recovery algorithms, such as OMP~\cite{tropp2004,Pati1993,davis1994}, CoSaMP~\cite{Tropp08}, or subspace pursuit~\cite{DM09}, can be modified to incorporate prior knowledge of the \emph{individual} sparsity levels of~\inputvec and/or~\error. Analytical recovery guarantees corresponding to the resulting modified algorithms do not seem to be available.}

\revision{We finally note that tightness of \fref{eq:P0_nothingknown_assump} can be established for $\dicta=\bF_M$ and $\dictb=\bI_M$.
Specifically, consider the pair $\inputvec=\comb{2\sqrt{\outputdim}}$, $\error=-\comb{2\sqrt{\outputdim}}$ and the  alternative pair $\inputvec'=\comb{2\sqrt{\outputdim}}-\comb{\sqrt{\outputdim}}$, $\error'=-\comb{2\sqrt{\outputdim}}+\comb{\sqrt{\outputdim}}$. 
It can be shown that both $\inputvec$ and $\inputvec'$ are in the admissible set of $(\Pzero,\,n_e)$ in~\eqref{eq:P0_nothinknown_problem}, satisfy $\normzero{\inputvec'}=\normzero{\inputvec}$,  and lead to the same measurement outcome \vecz. Therefore, (\Pzero, $n_e$) cannot distinguish between \inputvec and $\inputvec'$ (we refer to \cite{Kuppinger2011} for  details).}

\subsection{Case IV: No knowledge about the support sets}
\label{sec:noKnowledge}

Finally, we consider the case of no knowledge (prior to recovery) about the support sets \setX and~\setE. 
A corresponding application scenario would be the restoration of an audio signal (whose spectrum is sparse with unknown support set) that is corrupted by impulse noise, e.g., click or pop noise occurring at unknown locations. 
Another typical application can be found in the realm of signal separation; e.g.,  the decomposition of images into two distinct features, i.e., into a part that exhibits a sparse representation in the dictionary \dicta and another part that exhibits a sparse representation in \dictb.
Decomposition of the image \noiseout then amounts to performing sparse-signal recovery based on $\noiseout = \dicta\inputvec + \dictb\error$ with no knowledge about the support sets \setX and \setE available prior to recovery. 
The individual image features are given by $\dicta\inputvec$ and $\dictb\error$.

\sloppy

\revision{Recovery guarantees for this case follow from the results in~\cite{kuppinger2010a}. Specifically, by rewriting~\eqref{eq:systemmodel} as $\vecz=\dict\vecw$ as in \fref{eq:simpleconcatenation}, we can employ the recovery guarantees in~\cite{kuppinger2010a}, which are explicit in the coherence parameters \coha and \cohb, and the dictionary coherence \coh of \dict.}
\revision{For the sake of completeness, we restate the following result from \cite{kuppinger2010a}.}
\revision{\begin{thm}[\!\!{\cite[Thm.~2]{kuppinger2010a}}]\label{thm:spark_bound}
Let $\vecz=\dict\vecw$ with $\vecw=[\,\inputvec^T\,\,\error^T\,]^T$ and $\dict=\dictab$ with the coherence parameters $\coha\leq \cohb$ and the dictionary coherence $\coh$ as defined in \fref{eq:coherenceAB}. A sufficient condition for the vector \vecw to be the unique solution of (\Pzero) applied to $\noiseout=\dict\vecw$ is 
\begin{align}\label{eq:P0bound}
  \nx+\nerr=n_w < \frac{f(\xsol)+\xsol}{2}
\end{align}
where
\begin{align*}
f(x)= \dfrac{(1+\coha)(1+\cohb)-x\cohb(1+\coha)}{x(\coh^2-\coha\cohb)+\coha(1+\cohb)}
\end{align*}
and $\xsol=\min\{\xbord,\xstat\}$. Furthermore, $\xbord=(1+\cohb)/(\cohb+\coh^2)$ and 
\begin{align*}
\xstat=
\left\{\begin{array}{l}
1/ \coh, \,\, \text{if}\,\, \coha=\cohb=\coh, \\[0.3cm]
\dfrac{\coh\sqrt{(1+\coha)(1+\cohb)}-\coha-\coha\cohb}{\coh^2-\coha\cohb}, \,\, \text{otherwise.}
 \end{array}\right.		
\end{align*}
\end{thm}}
\revision{Obviously, once the vector \vecw has been recovered, we can extract~\inputvec and \vece. The following theorem, originally stated in \cite{kuppinger2010a}, guarantees that BP and OMP deliver the unique solution of (P0) applied to $\vecz=\dict\vecw$ and the associated recovery threshold, as shown in \cite{kuppinger2010a}, is only slightly more restrictive than that for (P0) in \fref{eq:P0bound}.}
\revision{\begin{thm}[\!\!{\cite[Cor.~4]{kuppinger2010a}}]\label{thm:l1_general}
A sufficient condition for BP and OMP  to deliver the unique solution of (P0) applied to $\noiseout=\dict\vecw$  is given by
\begin{align}\label{eq:l1_final}
n_w < 
\left\{\begin{array}{l}
\dfrac{\constoneA\big(\constoneB-(\coh+3\cohb)\big)}{2(\coh^2-\cohb^2)}, \,\, \text{if } \cohb<\coh  \text{ and }  \thlone >1,\vspace{0.2cm}\\[0.2cm]
\dfrac{1+2\coh^2 +3\cohb -\coh\constoneA}{2(\coh^2+\cohb)}, \,\, \text{otherwise}
\end{array}\right. 
\end{align}
with $n_w= \nx+\nerr $ and 
\begin{align*}
\thlone=\dfrac{\constoneA\sqrt{2\coh\left(\cohb+3\coh+\constoneB\right)}-2\coh - 2\cohb(\constoneA+\coh)}{2(\coh^2-\cohb^2)}
\end{align*}
where $\constoneA=1+\cohb$ and $\constoneB=2\sqrt{2}\sqrt{\coh(\cohb+\coh)}$.
\end{thm}
}

\fussy

We emphasize that both thresholds \fref{eq:P0bound} and \fref{eq:l1_final} are more restrictive than those in~\eqref{eq:P1_bothknown_assump}, \eqref{eq:P0uk_assump},~\eqref{eq:P1signaluk_assump}, and~\eqref{eq:P0_nothingknown_assump} (see also~\fref{sec:factortwoinguarantees}), which is consistent with the intuition that additional knowledge about the support sets \setX and \setE should lead to higher recovery thresholds.
\revision{Note that tightness of  \fref{eq:P0bound} and \fref{eq:l1_final} was established before in \cite{feuer2003} and  \cite{kuppinger2010a}, respectively.}



\section{Discussion of the Recovery Guarantees}\label{sec:discussion}
%
%
\revision{The aim of this section is to provide an interpretation of the recovery guarantees found in \fref{sec:main}. Specifically,  we discuss the impact of support-set knowledge on the recovery thresholds we found, and we point out  limitations of our results.}

\subsection{Factor of two in the recovery thresholds}
\label{sec:factortwoinguarantees}
Comparing the recovery thresholds~\eqref{eq:P1_bothknown_assump},~\eqref{eq:P0uk_assump},~\eqref{eq:P1signaluk_assump},~and~\eqref{eq:P0_nothingknown_assump} (Cases I--III), we observe that the price to be paid for not knowing the support set \setX or \setE is a reduction of the recovery threshold by a factor of two (note that in Case III, both \setX and \setE are unknown, but the cardinality of either \setX or \setE is known). For example, consider the recovery thresholds~\eqref{eq:P1_bothknown_assump} and~\eqref{eq:P0uk_assump}. For given $n_e\in[0,1+1/\cohb]$, solving~\eqref{eq:P1_bothknown_assump} for \nx yields
\be
	\nx<\frac{(1+\coha)(1-\cohb(n_e-1))}{n_e(\cohm^2-\coha\cohb)+\coha(1+\cohb)}.
\een
Similarly, still assuming $n_e\in[0,1+1/\cohb]$ and solving~\eqref{eq:P0uk_assump} for \nx, we get
\be
	\nx<\frac{1}{2}\,\frac{(1+\coha)(1-\cohb(n_e-1))}{n_e(\cohm^2-\coha\cohb)+\coha(1+\cohb)}.
\een
Hence, knowledge of \setX prior to recovery allows for the recovery of a signal with twice as many nonzero entries in \inputvec compared to the case where \setX is not known.
%
%
This factor-of-two penalty has the same roots  as the well-known factor-of-two penalty in spectrum-blind sampling~\cite{feng1996spectrum,Bresler2008,mishali2009a}. \revision{Note that the same factor-of-two penalty can be inferred from the RIC-based recovery guarantees in~\cite{Jacques2010,C08}, when comparing the recovery threshold specified in~\cite[Thm.~1]{Jacques2010} for signals where partial support-set knowledge is available (prior to recovery) to that given in \cite[Thm.~1.1]{C08} which does not assume prior support-set knowledge.}

\begin{figure}
\centering
 \includegraphics[width=0.95\columnwidth]{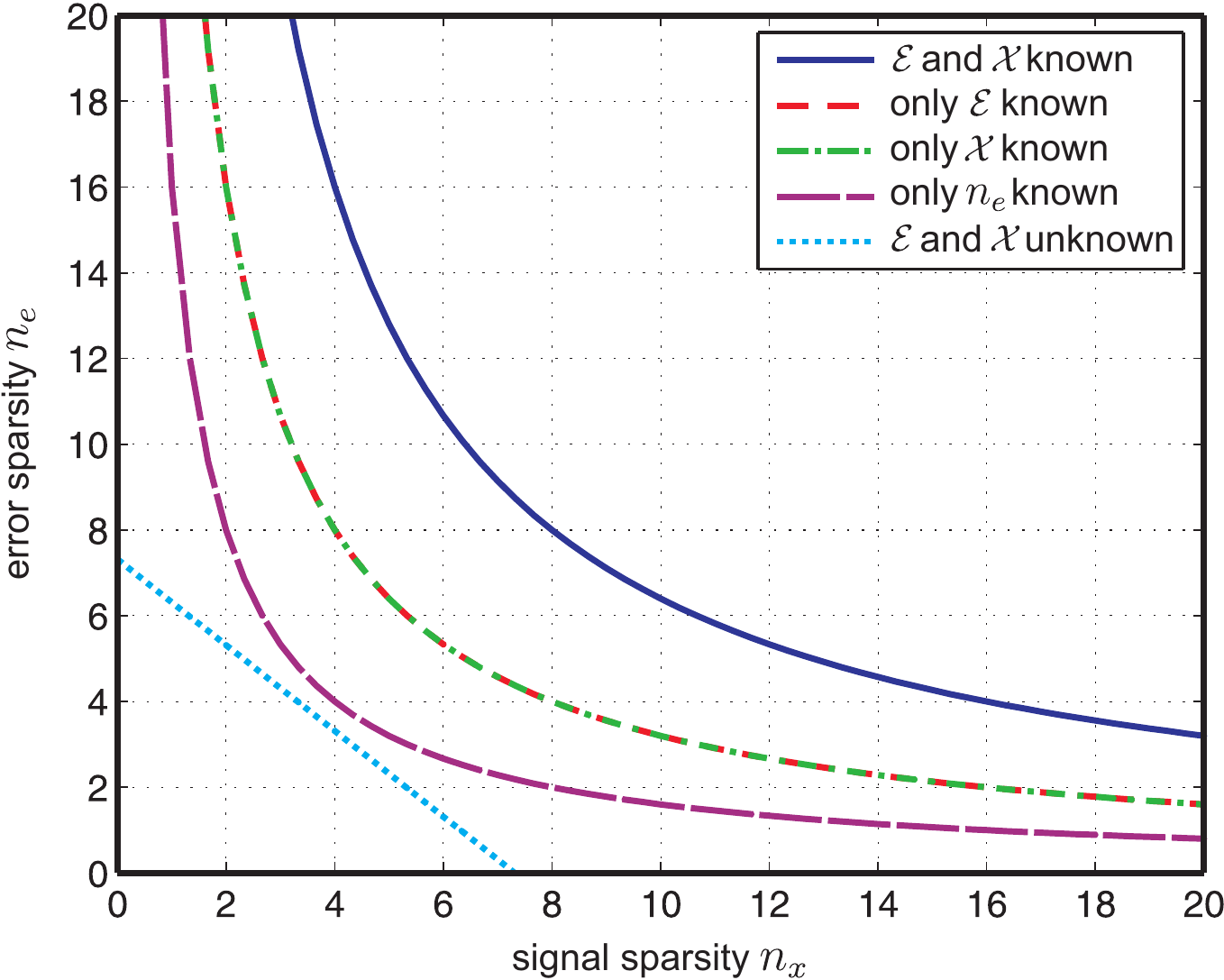}
  \caption{\revision{Recovery thresholds~\eqref{eq:P1_bothknown_assump},~\eqref{eq:P0uk_assump},~\eqref{eq:P1signaluk_assump},~\eqref{eq:P0_nothingknown_assump}, and \fref{eq:l1_final} for $\coha=\cohb=0$, and $\cohm=1/\sqrt{64}$. Note that the curves for ``only \setE known'' and ``only \setX known'' are on top of each other.}}
  \label{fig:th_th_ONB}
\end{figure}

\begin{figure}
\centering
 \includegraphics[width=0.95\columnwidth]{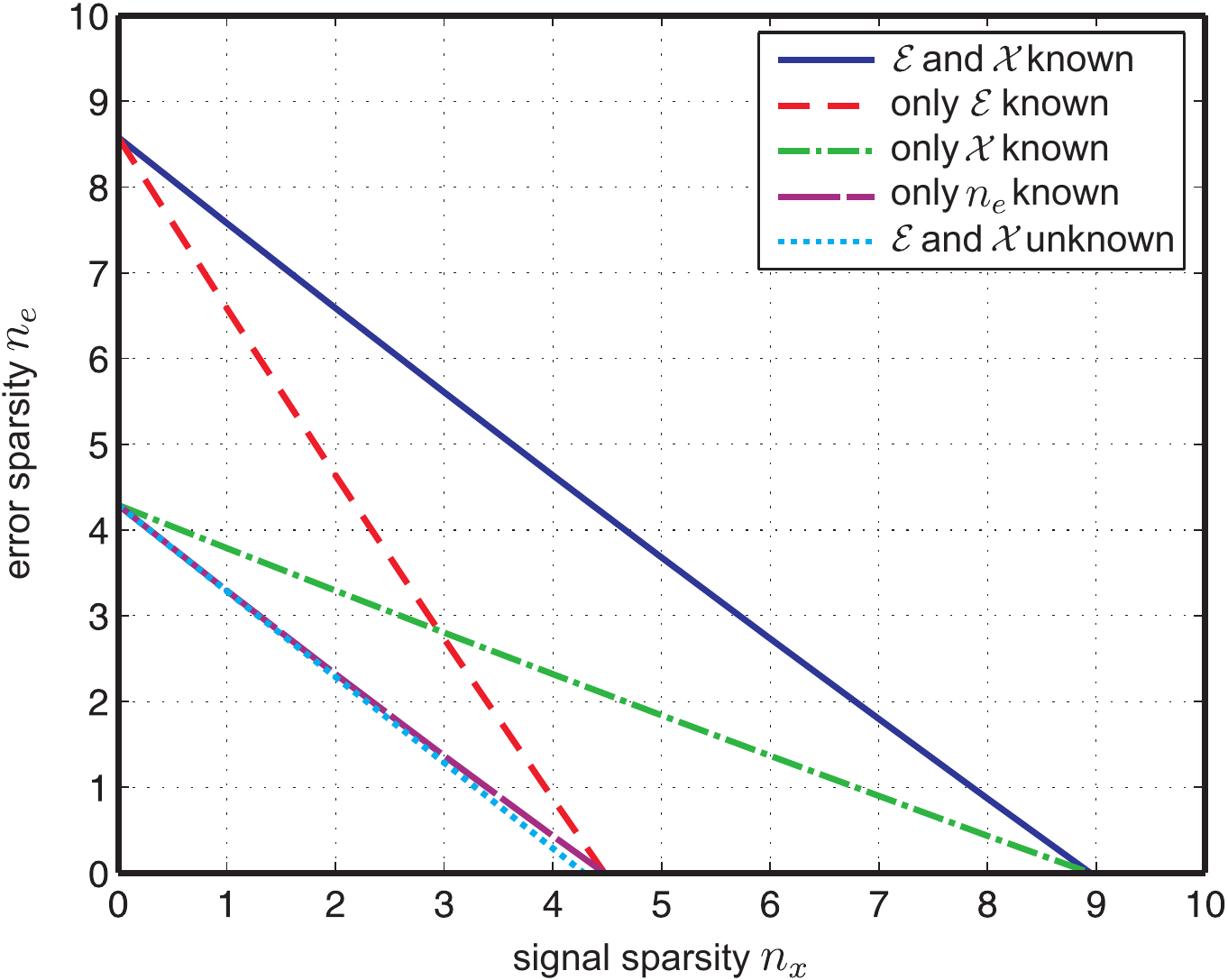}
  \caption{Recovery thresholds~\eqref{eq:P1_bothknown_assump},~\eqref{eq:P0uk_assump},~\eqref{eq:P1signaluk_assump},~\eqref{eq:P0_nothingknown_assump}, and~\fref{eq:l1_final} for $\coha = 0.1258$, $\cohb = 0.1319$, and $\cohm = 0.1321$.}
  \label{fig:th_th_sym}
\end{figure}

We illustrate the factor-of-two penalty in Figs.~\ref{fig:th_th_ONB} and~\ref{fig:th_th_sym}, where the recovery thresholds~\eqref{eq:P1_bothknown_assump},~\eqref{eq:P0uk_assump}, \eqref{eq:P1signaluk_assump}, \eqref{eq:P0_nothingknown_assump}, and \fref{eq:l1_final} 
are shown. 
In~\fref{fig:th_th_ONB}, we consider the case $\coha=\cohb=0$ and $\cohm=1/\sqrt{64}$. We can see that for \setX and \setE known the threshold evaluates to $\nx n_e<64$.
When only \setX or \setE is known we have $\nx n_e<32$, and finally in the case where only $n_e$ is known we get $\nx n_e<16$.
\revision{Note furthermore that in Case IV, where no knowledge about the support sets is available, the recovery threshold is more restrictive than in the case where $n_e$ is known.}

\revision{In~\fref{fig:th_th_sym}, we show the recovery thresholds for $\coha = 0.1258$, $\cohb = 0.1319$, and $\cohm = 0.1321$.
We see that all threshold curves are straight lines. 
This behavior can be explained by noting that (in contrast to the assumptions underlying \fref{fig:th_th_ONB}) the dictionaries \dicta and \dictb have $\coha,\cohb>0$ and the corresponding recovery thresholds are essentially dominated by the numerator of the RHS expressions in  \eqref{eq:P1_bothknown_assump},~\eqref{eq:P0uk_assump}, \eqref{eq:P1signaluk_assump}, and \fref{eq:P0_nothingknown_assump}, which depends on both \nx and \nerr. 
More concretely, if $\coha=\cohb=\cohm=\coh > 0$, then the recovery threshold for Case II (where the support set \setE is known) becomes
\begin{align} \label{eq:wowitreallyturnedouttobealinearthreshold}
 2\nx + \nerr < \Big(1+\coh^{-1}\Big)
\end{align}
which reflects the behavior observed in~\fref{fig:th_th_sym}. 
}


\sloppy

\subsection{The square-root bottleneck}

The  recovery thresholds presented in~\fref{sec:main} hold for \emph{all signal and noise realizations} \inputvec and \error and for \emph{all} dictionary pairs (with given coherence parameters).
\revision{However, as is well-known in the sparse-signal recovery literature, coherence-based recovery guarantees are---in contrast to RIC-based recovery guarantees---fundamentally limited by the so-called \emph{square-root bottleneck}~\cite{tropp2008}.} More specifically, in the noiseless case (i.e., for $\error=\mathbf{0}_{N_b}$), the threshold~\eqref{eq:classicalthreshold} states that recovery can be guaranteed only for up to $\sqrt{\outputdim}$ nonzero entries in \inputvec. 
Put differently, for a fixed number of nonzero entries \nx in \inputvec, i.e., for a fixed sparsity level, the number of measurements \outputdim required to recover \inputvec through (\Pzero), BP, or OMP is on the order of~$\nx^2$.

\revision{As in the classical sparse-signal recovery literature, the square-root bottleneck can be broken by performing a probabilistic analysis~\cite{tropp2008}.
This line of work---albeit interesting---is outside the scope of the present paper and is further investigated in \cite{kuppinger2010,kuppinger2010a,PBSB12}.}

\fussy

\subsection{Trade-off between \nx and \nerr}
\revision{We next illustrate a trade-off between the sparsity levels of \inputvec and
  \error.}  
\revision{Following the procedure outlined in~\cite{calderbank1997,gribonval2003}, we construct a dictionary \dicta consisting of $A$ ONBs and a dictionary \dictb consisting of $B$ ONBs such that $\mu_a = \mu_b = \mu_m= 1/\sqrt{\outputdim}$, where $A+B\le M+1$ with $M=p^k$, $p$  prime, and $k\in\naturals^+$. Now, let us assume that the error sparsity level scales according to $n_e = \alpha \sqrt{M}$ for some $0\le \alpha \le 1$.  For the case where  only $\setE$ is known but $\setX$ is unknown (Case II), we find from \fref{eq:P0uk_assump} that any signal $\inputvec$ with (order-wise) $(1-\alpha)\sqrt{M}/2$ non-zero entries (ignoring terms of order less than $\sqrt{\outputdim}$) can be reconstructed. Hence, there is 
a trade-off between the sparsity levels of $\inputvec$ and $\error$ (here quantified through the parameter $\alpha$), and both sparsity levels scale with~$\sqrt{M}$.}


%

\section{Numerical Results}
\label{sec:simulations}
%

We first report simulation results and  compare them to the corresponding analytical results in the paper. We will find that even though the analytical thresholds are pessimistic in general, they do reflect the numerically observed recovery behavior correctly. In particular, we will see that the factor-of-two penalty discussed in \fref{sec:factortwoinguarantees} can also be observed in the numerical results.
We then demonstrate, through a simple inpainting example, that perfect signal recovery in the presence of sparse errors is possible even if the corruptions are significant (in terms of the \elltwo-norm of the sparse noise vector \error).
\revision{In all numerical results, OMP  is  performed with a predetermined number of iterations~\cite{tropp2004,Pati1993,davis1994}, i.e., for Case II and Case IV, we set the number of iterations to $\nx$ and $\nx+\nerr$, respectively. To implement BP, we employ SPGL1~\cite{BF08,spgl1:2007}.}


\subsection{Impact of support-set knowledge on recovery thresholds}

We first compare simulation results to the recovery thresholds~\eqref{eq:P1_bothknown_assump}, \eqref{eq:P0uk_assump}, \eqref{eq:P1signaluk_assump}, and~\fref{eq:l1_final}. 
%
%
%
%
%
%
For a given pair of dictionaries \dicta and \dictb we generate signal vectors \inputvec and error vectors \error as follows:
We first fix \nx and $n_e$, then the support sets of the $\nx$-sparse vector \inputvec and the $n_e$-sparse vector \error are chosen uniformly at random among all possible  support sets of cardinality \nx and $n_e$, respectively. Once the support sets have been chosen, we generate the nonzero entries of \inputvec and \error by drawing from i.i.d.\ zero mean, unit variance Gaussian random variables.
%
%
%
For each pair of support-set cardinalities \nx and $n_e$, we perform 10\,000 Monte-Carlo trials 
and declare success of recovery whenever the recovered vector $\hat{\bmx}$ satisfies
\begin{align}\label{eq:success_def}
\normtwo{\hat{\inputvec}-\inputvec} <  10^{-3} \normtwo{\inputvec}.
\end{align}
%
We plot the 50\% success-rate contour, i.e., the border between the region of pairs (\nx, $n_e$) for which~\eqref{eq:success_def} is satisfied in at least 50\% of the trials and the region where~\eqref{eq:success_def} is satisfied in less than 50\% of the trials.
The recovered vector $\hat\inputvec$ is obtained as follows: 
\begin{itemize}
\item \emph{Case I:} When \setX and \setE are both known, we perform recovery according to~\eqref{eq:recoveryifeverythingisknown}. 
\item \emph{Case II:} When either only \setE or only \setX is known, we apply BP and OMP using the modified dictionary as detailed in \fref{thm:P1uk_equivalence} and~\fref{cor:P1_signal_known}, respectively.
\item \emph{Case IV:} \revision{When neither \setX nor \setE is known, we apply BP and OMP to the concatenated dictionary $\dict=[\,\dicta\,\,\dictb\,]$ as described in \fref{thm:l1_general}.}
\end{itemize}
\revision{Note that for \emph{Case III}, i.e., the case where the cardinality $n_e$ of the support set \setE is known---as pointed out in~\fref{sec:cardinalityKnown}---we only have uniqueness results but no analytical recovery guarantees, neither for BP nor for greedy recovery algorithms that make use of the \emph{separate} knowledge of $n_x$ or $n_e$ (whereas, e.g., standard OMP makes use of knowledge of $n_x+n_e$, rather than knowledge of $n_x$ and $n_e$ individually).
This case is, therefore, not considered in the simulation results below.}

\begin{figure}[t]
  \centering
  \includegraphics[width=0.95\columnwidth]{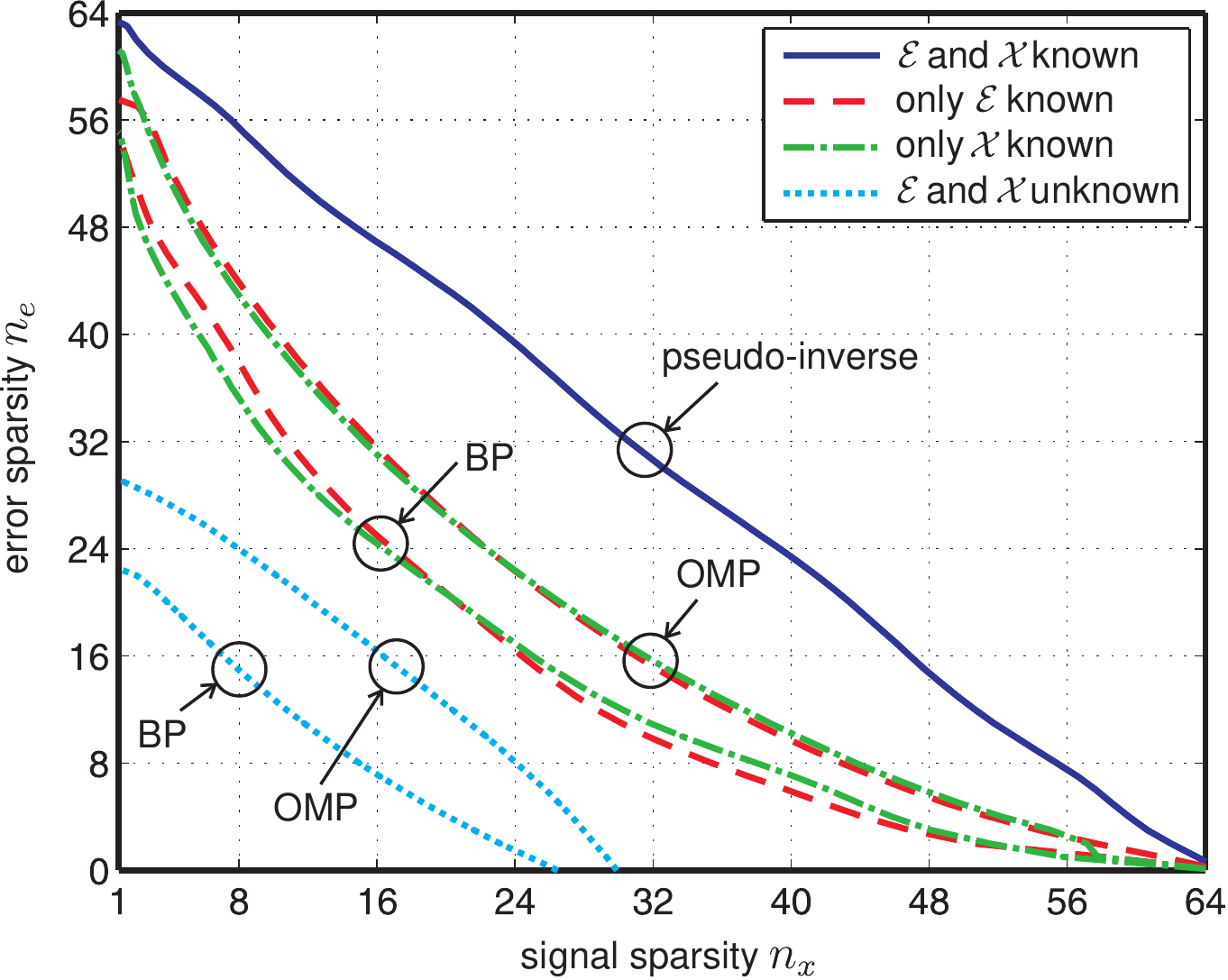}
  \caption{Impact of support-set knowledge on the 50\% success-rate contour of OMP and BP for the Hadamard--identity pair.}
  \label{fig:pt_bpvsomp}
\end{figure}

\begin{figure}[t]
  \centering
  \includegraphics[width=0.95\columnwidth]{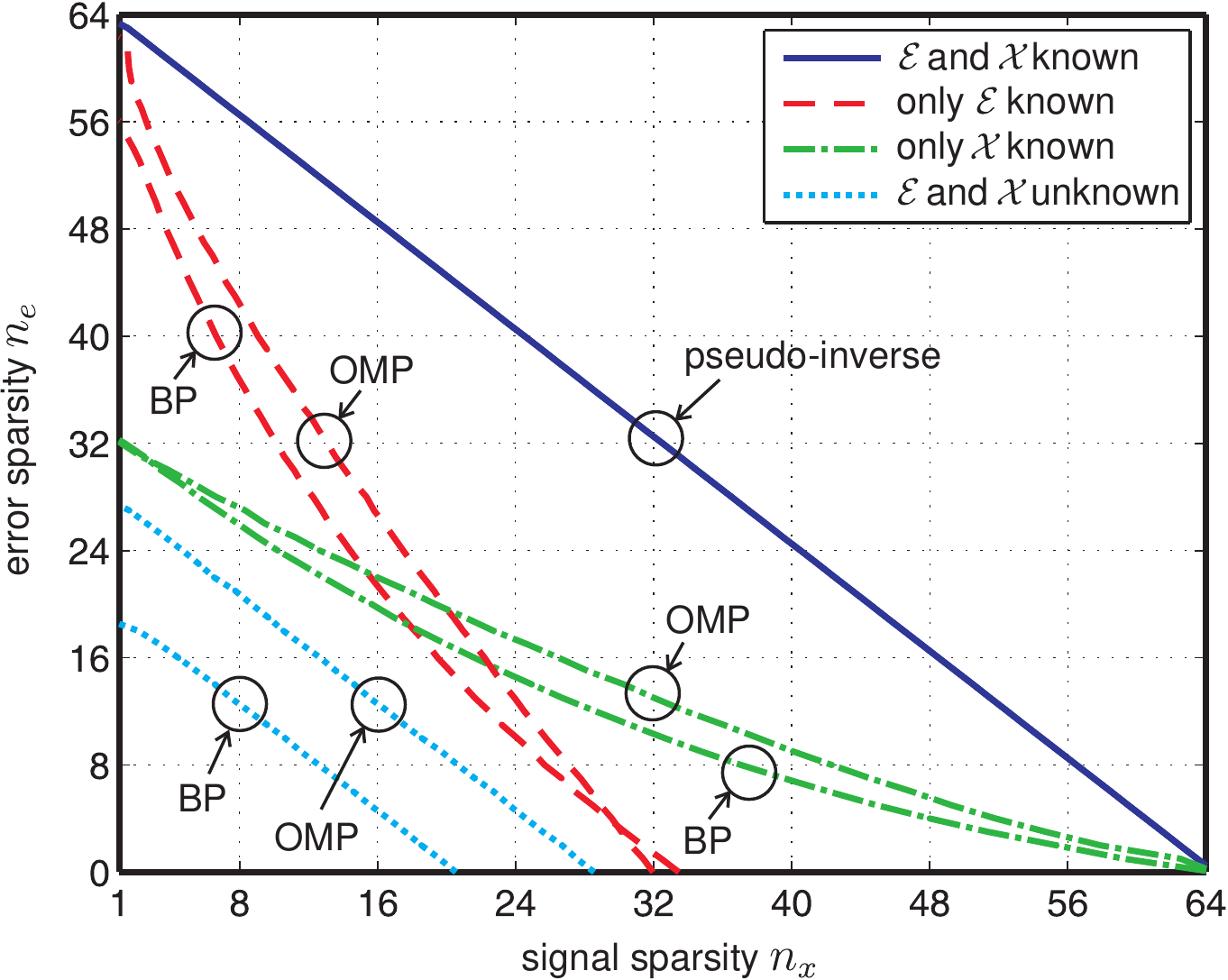}
  \caption{Impact of support-set knowledge on the 50\% success-rate contour of OMP and BP performing recovery in pairs of  approximate ETFs each of dimension $64\times80$.}
  \label{fig:pt_orthvsrand_OMP}
\end{figure}

\subsubsection{Recovery performance for the Hadamard--identity pair using BP and OMP}  

\label{sec:AB_ONBs}
We take $\outputdim=64$, let \dicta be the Hadamard ONB~\cite{agaian1985}  and set $\dictb=\bI_\outputdim$, which results in $\coha=\cohb=0$ and $\cohm=1/\sqrt{\outputdim}$.
\fref{fig:pt_bpvsomp} shows  50\% success-rate contours, under different assumptions of support-set knowledge. 
For perfect knowledge of \setX and \setE, we observe that the 50\% success-rate contour is at about $\nx+n_e\approx\outputdim$, which is significantly better than the sufficient condition $\nx n_e < \outputdim$ (guaranteeing perfect recovery) provided in~\eqref{eq:P1_bothknown_assump}.\footnote{For $\dicta=\bF_\outputdim$ and $\dictb=\bI_\outputdim$  it was proven in~\cite{tropp2008linear} that a set of columns  chosen randomly from both \dicta and \dictb is linearly independent (with high probability) given that the total number of chosen columns, i.e., $\nx+n_e$ here, does not exceed a constant proportion of \outputdim.}  
When either only $\setX$ or only $\setE$ is known, the recovery performance is essentially independent of whether $\setX$ or \setE is known. This is also reflected by the analytical thresholds~\eqref{eq:P0uk_assump} and~\eqref{eq:P1signaluk_assump} when evaluated for $\coha=\cohb=0$ (see also~\fref{fig:th_th_ONB}). Furthermore, OMP is seen to outperform BP. 
When neither \setX nor \setE is known, OMP again outperforms BP.

%
It is interesting to see that the factor-of-two penalty discussed in~\fref{sec:factortwoinguarantees} is reflected in \fref{fig:pt_bpvsomp} (for $\nx=n_e$) between Cases I and II. 
Specifically, we can observe that for full support-set knowledge (Case I)  the 50\% success-rate is achieved at $\nx=n_e\approx31$. If either \setX or \setE only is known (Case II), OMP achieves 50\% success-rate at $\nx=n_e\approx23$, demonstrating a factor-of-two penalty since $31\cdot31\approx23\cdot23\cdot2$. 
Note that the results from BP in \fref{fig:pt_bpvsomp} do not seem to reflect the factor-of-two penalty. 
For lack of an efficient recovery algorithm (making use of knowledge of $n_e$) we do not show numerical results for Case III. 
\subsubsection{Impact of $\coha,\cohb>0$} 

We take $\outputdim=64$ and generate the dictionaries \dicta and \dictb as follows. Using the  alternating projection method described in~\cite{tropp2005}, we generate an approximate equiangular tight frame (ETF) for $\reals^\outputdim$ consisting of $160$ columns. We split this frame into two sets of $80$ elements (columns) each and organize them in the matrices \dicta and \dictb such that the corresponding coherence parameters are given by $\coha \approx 0.1258$, $\cohb \approx 0.1319$, and $\cohm \approx 0.1321$.
\fref{fig:pt_orthvsrand_OMP}  shows the 50\% success-rate contour under four different assumptions of support-set knowledge. In the case where either only \setX or only \setE is known and in the case where \setX and \setE are unknown, we use OMP and BP for recovery.
\revision{It is interesting to note that the graphs for the cases where only \setX or only \setE are known, are symmetric with respect to the line $\nx=\nerr$. This symmetry is also reflected in the analytical thresholds~\eqref{eq:P0uk_assump} and~\eqref{eq:P1signaluk_assump}  (see also~\fref{fig:th_th_sym} and the discussion in \fref{sec:factortwoinguarantees}).}

\revision{We finally note that in all cases considered above, the numerical results show that recovery is possible for significantly higher sparsity levels $\nx$ and $n_e$ than indicated by the corresponding analytical thresholds~\eqref{eq:P1_bothknown_assump}, \eqref{eq:P0uk_assump}, \eqref{eq:P1signaluk_assump}, and~\fref{eq:l1_final} (see also Figs.~\ref{fig:th_th_ONB} and~\ref{fig:th_th_sym}).}
The underlying reasons are 
\begin{inparaenum}[i)]
\item  the deterministic nature of the results, i.e., the recovery guarantees in~\eqref{eq:P1_bothknown_assump},~\eqref{eq:P0uk_assump},~\eqref{eq:P1signaluk_assump}, and~\fref{eq:l1_final} are valid for \emph{all} dictionary pairs (with given coherence parameters) and \emph{all} signal and noise realizations (with given sparsity level), and 
\item we plot the 50\% success-rate contour, whereas the analytical results guarantee \emph{perfect} recovery in 100\% of the cases.
\end{inparaenum}

\sloppy

\subsection{Inpainting example}
In transform coding one is typically interested in maximally sparse representations of a given signal to be encoded~\cite{jayant1984}. In our setting, this would mean that the dictionary \dicta should be chosen so that it leads to maximally sparse representations of a given family of signals. We next demonstrate, however, that in the presence of structured noise, the signal dictionary \dicta should \emph{additionally} be incoherent to the noise dictionary \dictb. This extra requirement can lead to very different criteria for designing transform bases (frames).

\fussy

\begin{figure*}[t]
\centering
\subfigure[Corrupted  image \newline ($\textsf{MSE}=-11.2$\,dB)]{
\includegraphics[width=0.47\columnwidth]{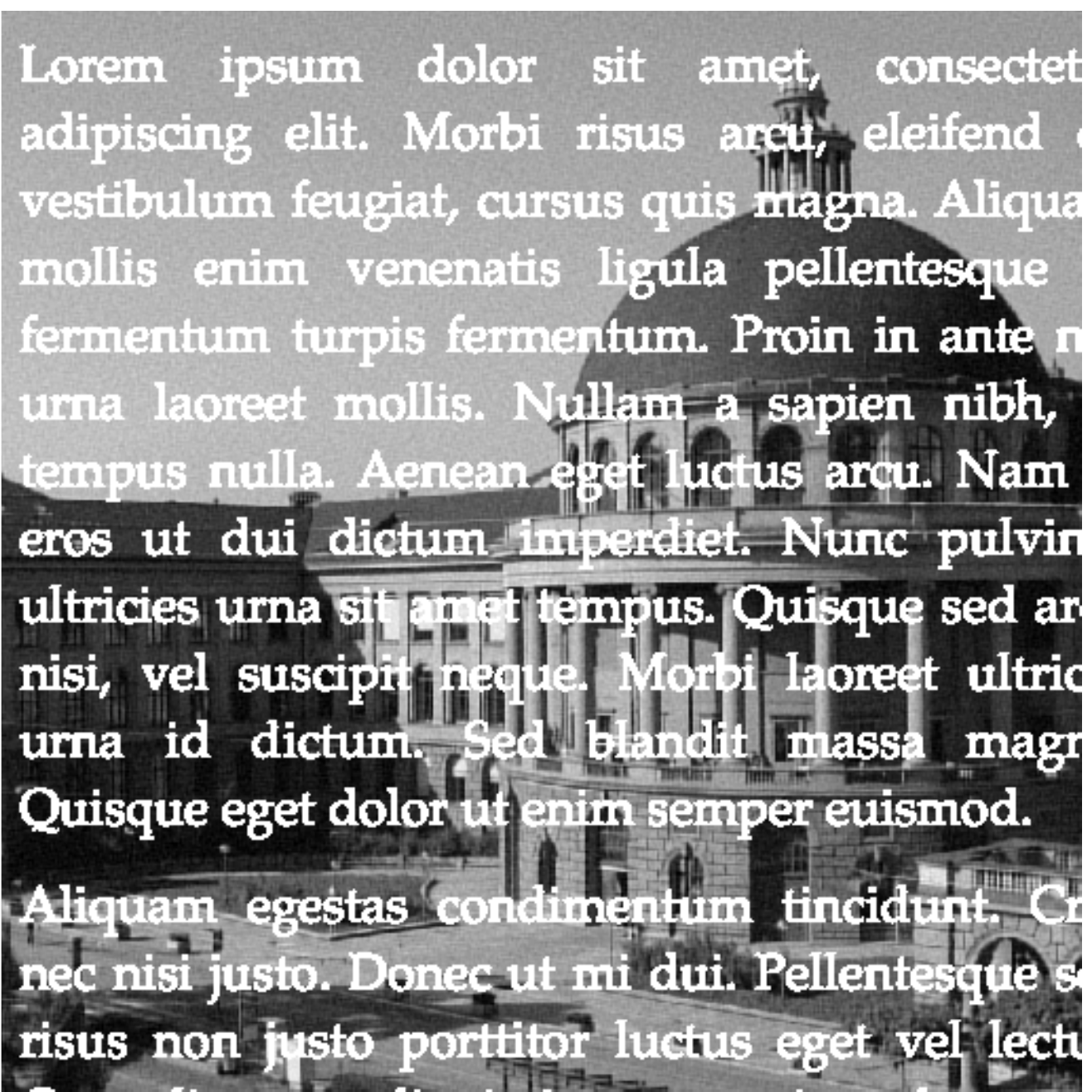}
\label{fig:dct_subfig1}
}
\subfigure[Recovery when $\setX$ and $\setE$  are known ($\textsf{MSE}=-184.6$\,dB)]{
\includegraphics[width=0.47\columnwidth]{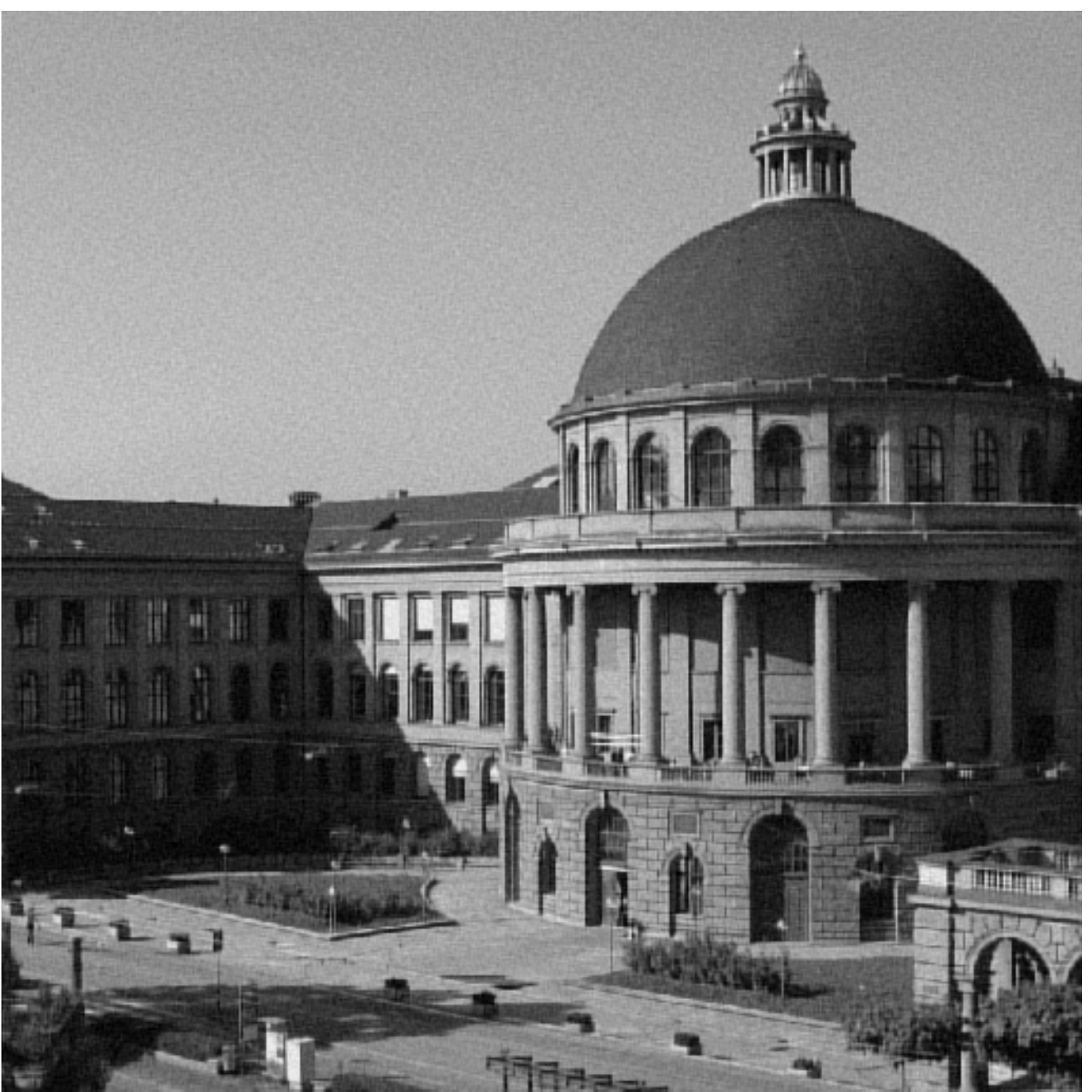}
\label{fig:dct_subfig2}
}
\subfigure[Recovery when only $\setE$ is known ($\textsf{MSE}=-113.3$\,dB)]{
\includegraphics[width=0.47\columnwidth]{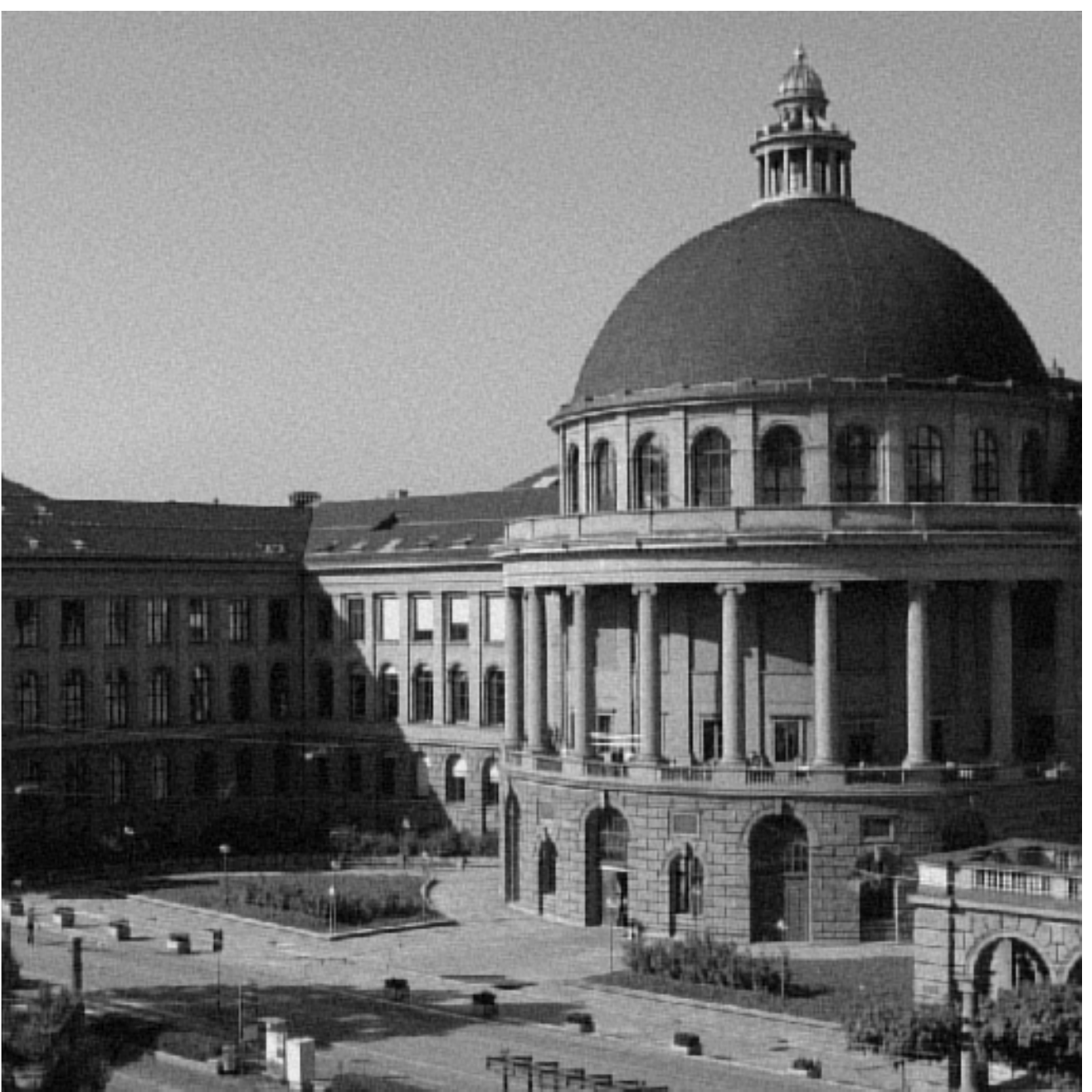}
\label{fig:dct_subfig3}
}
\subfigure[Recovery for \setX and \setE unknown ($\textsf{MSE}=-13.0$\,dB)]{
\includegraphics[width=0.47\columnwidth]{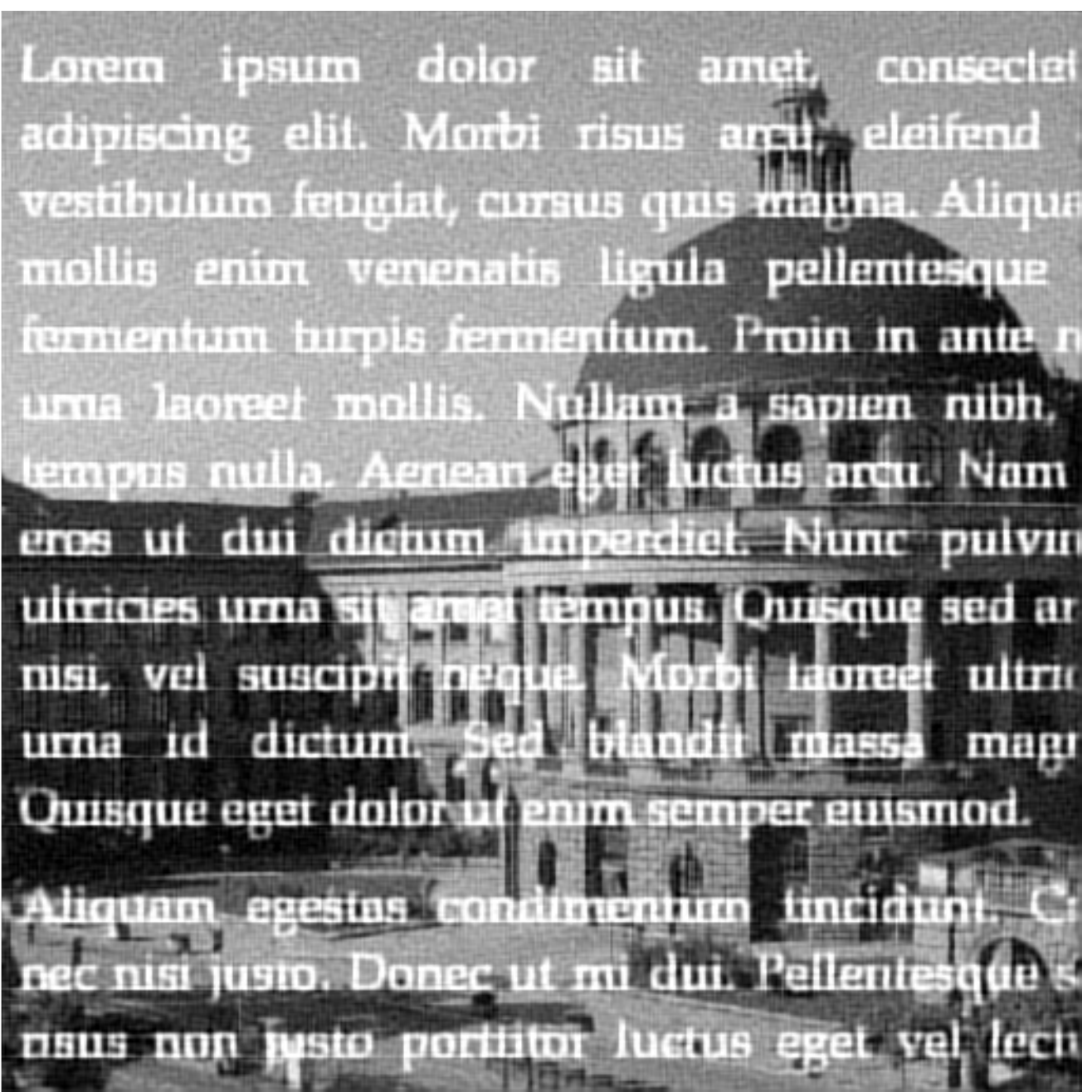}
\label{fig:dct_subfig4}
}
\caption{Recovery results using the DCT basis for the signal dictionary and the identity basis for the noise dictionary,  for the cases where \subref{fig:wavelet_subfig2} $\setX$ and $\setE$ are known, \subref{fig:wavelet_subfig3} only $\setE$ is known, and  \subref{fig:wavelet_subfig4}  no support-set knowledge is available. (Picture origin: ETH Z\"urich/Esther Ramseier).}
\label{fig:dct_recovery}
\vspace{0.4cm}
\end{figure*}

\begin{figure*}[t]
\centering
\subfigure[{Corrupted image \newline ($\textsf{MSE}=-11.2$\,dB)}]{
\includegraphics[width=0.47\columnwidth]{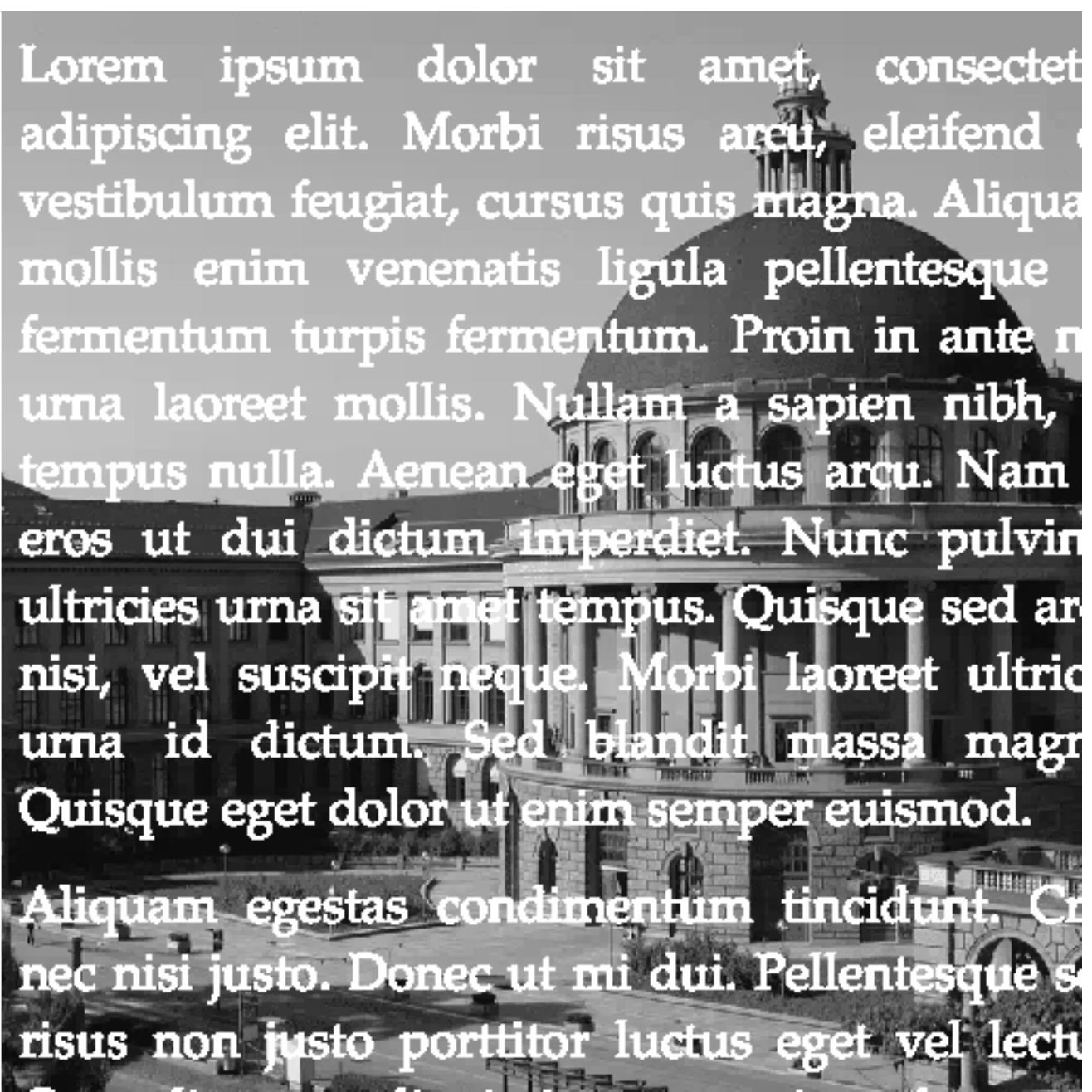}
\label{fig:wavelet_subfig1}
}
\subfigure[Recovery when $\setX$ and $\setE$ are known ($\textsf{MSE}=-27.1$\,dB)]{
\includegraphics[width=0.47\columnwidth]{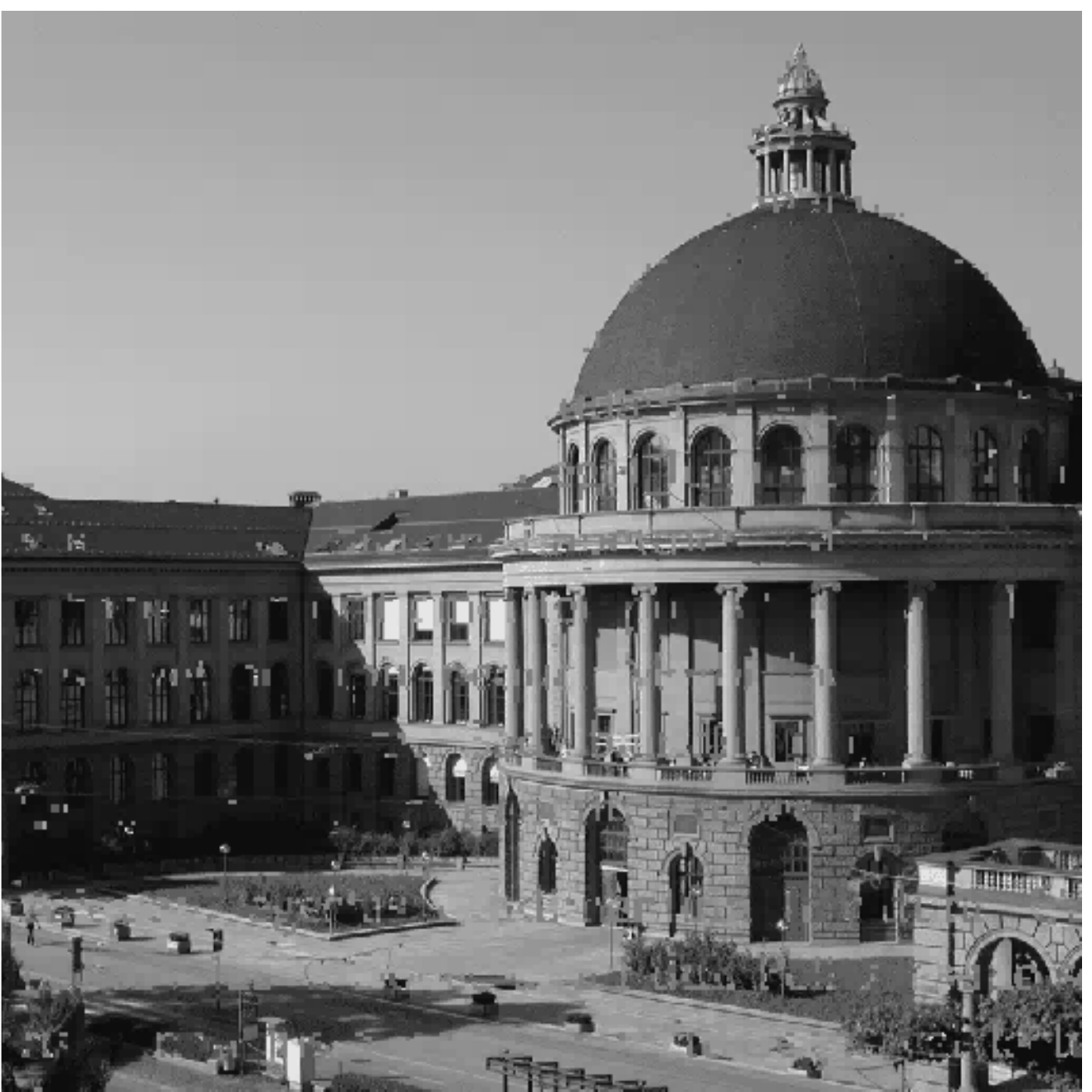}
\label{fig:wavelet_subfig2}
}
\subfigure[Recovery when only $\setE$ is known ($\textsf{MSE}=-27.1$\,dB)]{
\includegraphics[width=0.47\columnwidth]{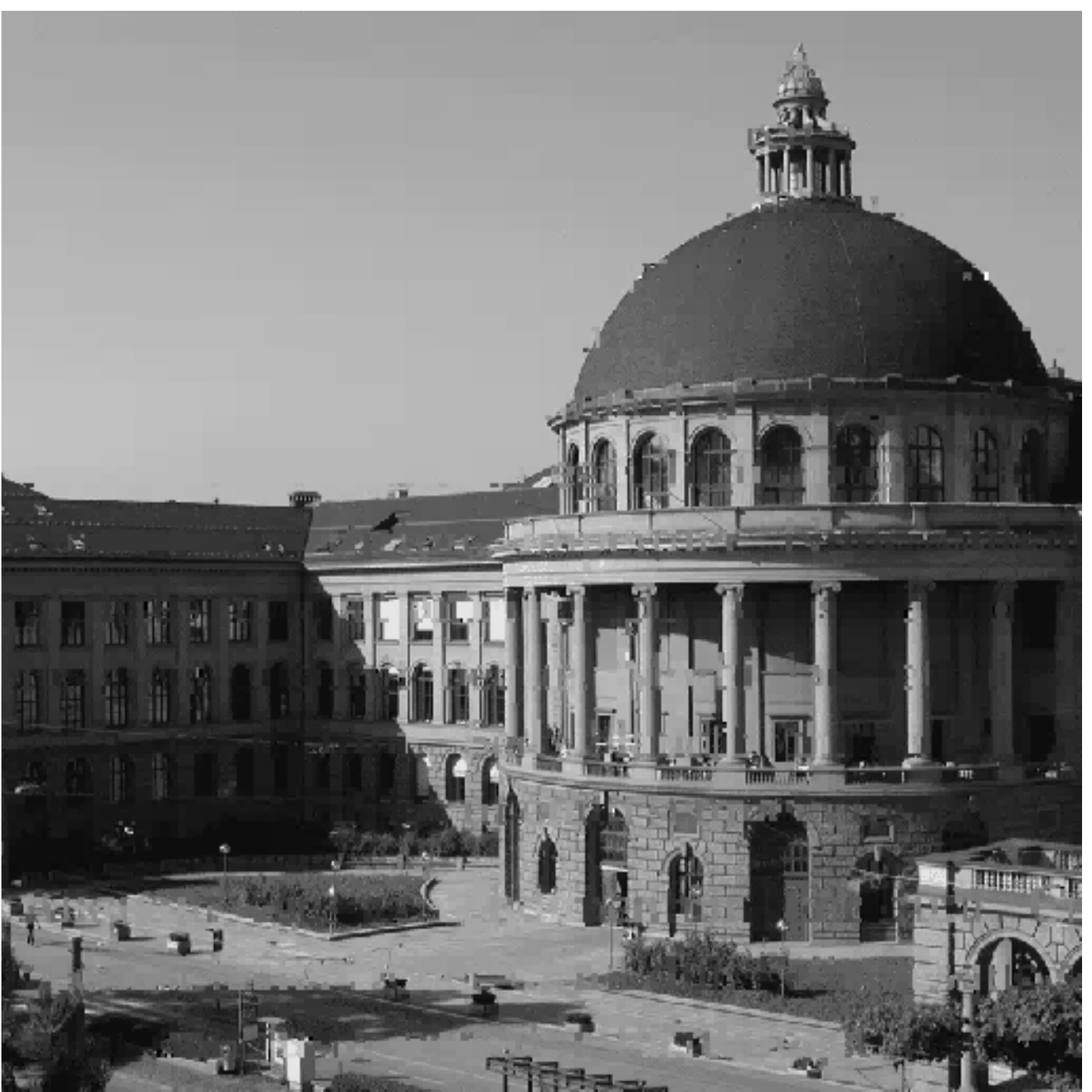}
\label{fig:wavelet_subfig3}
}
\subfigure[Recovery for \setX and \setE unknown ($\textsf{MSE}=-12.0$\,dB)]{
\includegraphics[width=0.47\columnwidth]{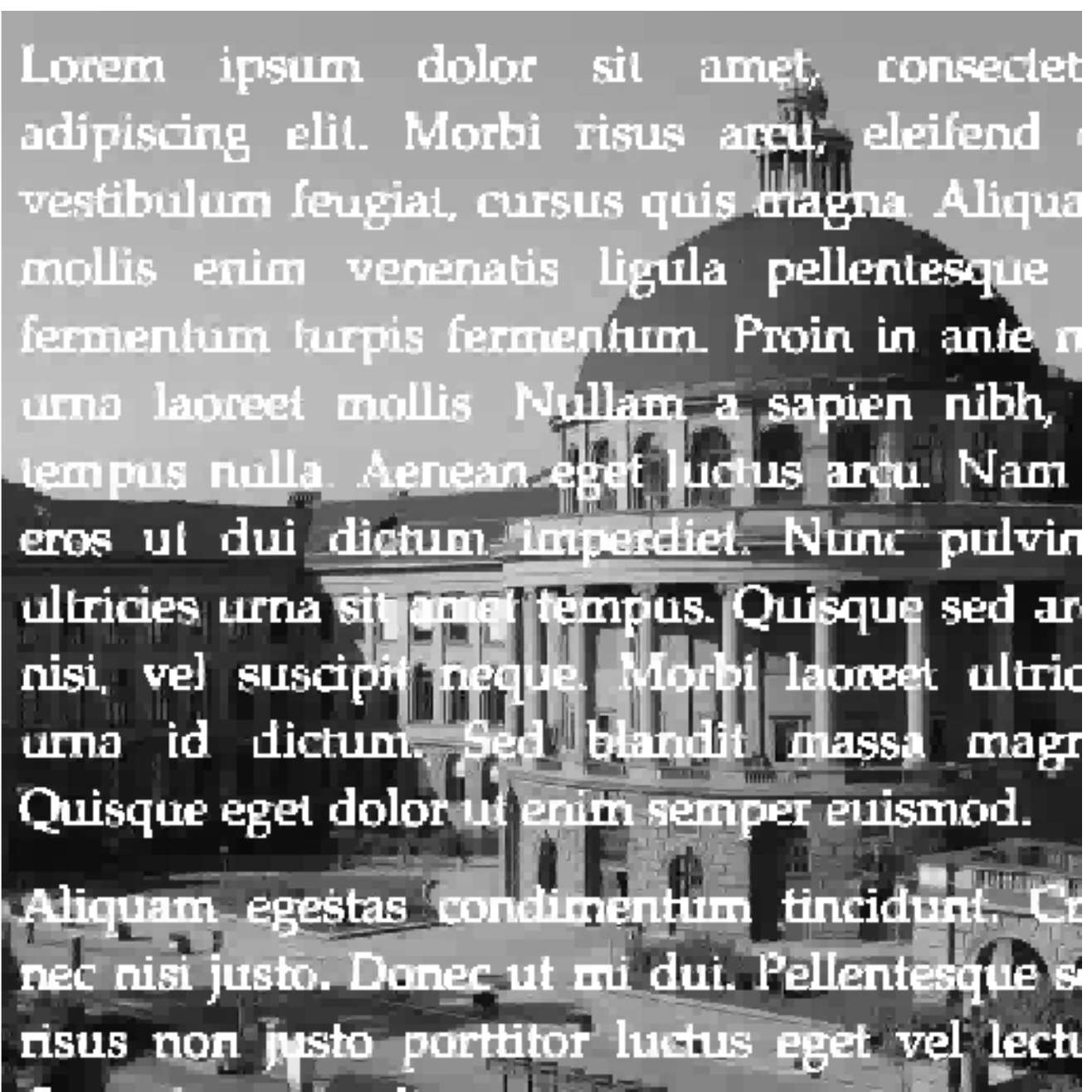}
\label{fig:wavelet_subfig4}
}
\caption{Recovery results for the signal dictionary given by the Haar wavelet basis and  the noise dictionary given by the identity basis, for the cases where \subref{fig:wavelet_subfig2} both $\setX$ and $\setE$ are known, \subref{fig:wavelet_subfig3} only $\setE$ is known, and  \subref{fig:wavelet_subfig4} no support-set knowledge is available. (Picture origin: ETH Z\"urich/Esther Ramseier).} 
\label{fig:wavelet_recovery}
\vspace{0.11cm}
\end{figure*}

To illustrate this point, and to show that perfect recovery can be guaranteed even when the $\elltwo$-norm of the noise term $\dictb\error$ is large, we consider the recovery of a sparsely corrupted 512$\times$512-pixel grayscale image of the main building of ETH Zurich. The dictionary \dicta is taken to be either the two-dimensional discrete cosine transform (DCT) or the Haar wavelet decomposed on three octaves~\cite{mallat1999wavelet}.
We first ``sparsify'' the image by retaining the 15\% largest entries of the image's representation \inputvec in \dicta.
%
%
%
We then corrupt (by overwriting with text) 18.8\% of the pixels  in the sparsified image by setting them to the brightest grayscale value; this means that the errors are sparse in~$\dictb=\bI_\outputdim$ and that the noise is structured but may have large $\elltwo$-norm. Image recovery is performed according to \fref{eq:recoveryifeverythingisknown} if $\setX$ and $\setE$ are known.
(Note, however, that knowledge of \setX is usually not available in inpainting applications.)
We use BP when only $\setE$ is known and when neither $\setX$ nor $\setE$ are known. 
The recovery results are evaluated by computing the mean-square error (MSE) between the sparsified image and its recovered version.
%
%

%
Figs.~\ref{fig:dct_recovery} and~\ref{fig:wavelet_recovery} show the corresponding results. As expected, the MSE increases as the amount of knowledge about the support sets decreases.
%
%
More interestingly, we note that even though Haar wavelets often yield a smaller approximation error in classical transform coding compared to the DCT, here the wavelet transform performs worse than the DCT.
%
%
This behavior is due to the fact that  sparsity is not the only factor determining the performance of a transform coding basis (or frame) in the presence of structured noise.
Rather the mutual coherence between the dictionary used to represent the signal and that used to represent the structured noise becomes highly relevant.
Specifically, in the example at hand, we have $\cohm=1/2$ for the Haar-wavelet and the identity basis, and $\cohm\approx0.004$ for the DCT and the identity basis.
%
%
%
The dependence of the analytical thresholds~\eqref{eq:P1_bothknown_assump},~\eqref{eq:P0uk_assump},~\eqref{eq:P1signaluk_assump},~\eqref{eq:P0_nothingknown_assump}, and~\fref{eq:l1_final} on the mutual coherence $\cohm$ explains the performance difference between the Haar wavelet basis and the DCT basis. 
%
\revision{An intuitive explanation for this behavior is as follows: The Haar-wavelet basis contains only four non-zero entries in the columns associated to fine scales, which is reflected in the high mutual coherence (i.e., $\cohm=1/2$) between the Haar-wavelet basis and the identity basis.
Thus, when projecting onto the orthogonal complement of $(\bI_M)_{\setE}$, it is likely that all non-zero entries of such columns are deleted, resulting in columns of all zeros.  Recovery of the corresponding non-zero entries of \inputvec is thus not possible.
%
}
%
%
%
%
In summary, we see that the choice of the transform basis (frame) for a sparsely corrupted signal  should 
not only aim at sparsifying the signal as much as possible but should also take into account the mutual coherence between the transform basis (frame) and the  noise sparsity basis (frame).
%

\section{Conclusion}\label{sec:conclusion}

%

%

The setup considered in this paper, in its generality, appears to be new and a number of interesting extensions are possible.
%
\revision{
In particular, developing (coherence-based) recovery guarantees for greedy algorithms such as CoSaMP~\cite{Tropp08} or subspace pursuit~\cite{DM09} for all cases studied in the paper are interesting open problems.}
%
\revision{Note that probabilistic recovery guarantees for the case where nothing is known about the signal and noise support sets (i.e., Case IV) readily follow from the results in~\cite{kuppinger2010a}. Probabilistic recovery guarantees for the other cases studied in this paper are in preparation~\cite{PBSB12}. 
Furthermore, an extension of the results in this paper that accounts for measurement noise (in addition to sparse noise) and applies to approximately sparse signals can be found in \cite{SB11}.}

\section*{Acknowledgments}

\revision{The authors would like to thank C.~Aubel, A.~Bracher, and G.~Durisi for interesting discussions, and the anonymous reviewers for their valuable comments which helped to improve the exposition.}

\appendices

\section{Proof of~\fref{thm:P1_bothknown}}
\label{app:P1_bothknown_proof}
We prove the full (column-)rank property of $\dict_{\setX,\setE}$ by showing that under~\eqref{eq:P1_bothknown_assump} there is a unique pair $(\inputvec,\error)$ with $\supp(\inputvec)=\setX$ and $\supp(\error)=\setE$ satisfying $\noiseout=\dicta\inputvec+\dictb\error$. Assume that there exists an alternative pair $(\inputvec',\error')$ such that $\noiseout=\dicta\inputvec'+\dictb\error'$ with $\supp(\inputvec')\subseteq\setX$ and $\supp(\error')\subseteq\setE$ (i.e., the support sets of $\inputvec'$ and $\error'$ are contained in \setX and \setE, respectively).
This would then imply that  
\begin{align*}
 \dicta\inputvec+\dictb\error=\dicta\inputvec'+\dictb\error'
\end{align*}
and thus
\begin{align*}
  \dicta(\inputvec-\inputvec')=\dictb(\error'-\error).
\end{align*}
Since both \inputvec and $\inputvec'$ have support in \setX it follows that $\inputvec-\inputvec'$ also has support in \setX, which implies $\normzero{\inputvec-\inputvec'}\leq \nx $. Similarly, we get $\normzero{\error'-\error}\leq n_e$.  Defining $\vecp=\inputvec-\inputvec'$ and $\setP=\supp(\inputvec-\inputvec')\subseteq\setX$, and, similarly, $\vecq=\error'-\error$ and $\setQ=\supp(\error'-\error)\subseteq\setE$, we obtain the following chain of inequalities:
\begin{align}
\nx n_e&\geq\normzero{\vecp}\normzero{\vecq}\nonumber = \abs{\setP}\abs{\setQ}\nonumber\\
 &\geq \frac{\pos{1-\coha(\abs{\setP}-1)} \pos{1-\cohb(\abs{\setQ}-1)}}{\cohm^2}\label{eq:bothknown1} \\ 
 & \geq \frac{\pos{1-\coha(\nx-1)} \pos{1-\cohb(n_e-1)}}{\cohm^2} = f(\nx,n_e)\label{eq:bothknown2}
\end{align}
where~\eqref{eq:bothknown1} follows by applying the uncertainty relation in~\fref{thm:uncertainty} (with $\epsilon_\setP=\epsilon_\setQ=0$ since both \vecp and \vecq are perfectly concentrated to \setP and \setQ, respectively) and~\eqref{eq:bothknown2} is a consequence of $\abs{\setP}\leq\nx$ and $\abs{\setQ}\leq n_e$. Obviously,~\eqref{eq:bothknown2} contradicts the assumption in~\eqref{eq:P1_bothknown_assump}, which completes the proof.
%
%

\section{Proof of~\fref{thm:P0_uk}}
\label{app:P0P1_uk_proof}
We begin by proving that \inputvec is the unique solution of $(\Pzero,\setE)$ applied to $\vecz=\dicta\inputvec+\dictb\error$. 
Assume that there exists an alternative vector $\inputvec'$ that satisfies $\dicta\inputvec'\in(\{\vecz\}+\setR(\dictb_\setE))$ with $\normzero{\inputvec'}\leq\nx$. This would imply the existence of a vector $\vece'$ with $\supp(\error')\subseteq\setE$, such that
%
%
\begin{align*}
 \dicta\inputvec+\dictb\error=\dicta\inputvec'+\dictb\error'
\end{align*}
and hence
\begin{align*}
  \dicta(\inputvec-\inputvec')=\dictb(\error'-\error).
\end{align*}
Since $\supp(\error)=\setE$ and $\supp(\error')\subseteq\setE$, we have $\supp(\error'-\error)\subseteq\setE$ and hence $\normzero{\error'-\error}\leq n_e$. Furthermore, since both \inputvec and $\inputvec'$ have at most \nx nonzero entries, we have $\normzero{\inputvec-\inputvec'}\leq2\nx$. Defining $\vecp=\inputvec-\inputvec'$ and $\setP=\supp(\inputvec-\inputvec')$, and, similarly, $\vecq=\error'-\error$ and $\setQ=\supp(\error'-\error)\subseteq\setE$, we obtain the following chain of inequalities
\begin{align}
2\nx n_e&\geq\normzero{\vecp}\normzero{\vecq} = \abs{\setP}\abs{\setQ}\nonumber\\
 & \geq \frac{\pos{1-\coha(\abs{\setP}\!-\!1)} \pos{1-\cohb(\abs{\setQ}\!-\!1)}}{\cohm^2}\label{eq:Xunknown1} \\ 
 & \geq \frac{\pos{1-\coha(2\nx\!-\!1)} \pos{1-\cohb(n_e\!-\!1)}}{\cohm^2}=f(2\nx,n_e)\label{eq:Xunknown2}
\end{align}
where~\eqref{eq:Xunknown1} follows by applying the uncertainty relation in~\fref{thm:uncertainty}  (with $\epsilon_\setP=\epsilon_\setQ=0$ since both \vecp and \vecq are perfectly concentrated to \setP and \setQ, respectively) and~\eqref{eq:Xunknown2} is a consequence of  $\abs{\setP}\leq2\nx$ and $\abs{\setQ}\leq n_e$. Obviously,~\eqref{eq:Xunknown2} contradicts the assumption in~\eqref{eq:P0uk_assump}, which concludes the first part of the proof. 


We next prove that \inputvec is also the unique solution of $(\textrm{BP},\setE)$ applied to $\vecz=\dicta\inputvec+\dictb\error$. 
Assume that there exists an alternative vector $\inputvec'$ that satisfies $\dicta\inputvec'\in(\{\vecz\}+\setR(\dictb_\setE))$ with $\normone{\inputvec'}\leq\normone{\inputvec}$. This would imply the existence of a vector $\vece'$ with $\supp(\error')\subseteq\setE$, such that
%
%
%
\begin{align*}
	\dicta\inputvec+\dictb\error=\dicta\inputvec'+\dictb\error'
\end{align*}
and hence
\begin{align*}
\dicta(\inputvec-\inputvec')=\dictb(\error'-\error).
\end{align*}
Defining $\vecp=\inputvec-\inputvec'$, we obtain the following lower bound for the \ellone-norm of $\inputvec'$ 
\begin{align}
\normone{\inputvec'}  = \normone{\inputvec-\vecp} & = \normone{\bP_\setX(\inputvec-\vecp)}+\normone{\bP_{\setX^c}\vecp}\nonumber\\
& \geq \normone{\bP_\setX\inputvec}-\normone{\bP_\setX\vecp}+\normone{\bP_{\setX^c}\vecp}\label{eq:XunknownBP1} \\
& = \normone{\inputvec}-\normone{\bP_\setX\vecp}+\normone{\bP_{\setX^c}\vecp}\nonumber
\end{align}
where~\eqref{eq:XunknownBP1} is a consequence of the reverse triangle inequality. Now, the \ellone-norm of $\inputvec'$ can be smaller than or equal to  that of \inputvec only if $\normone{\bP_\setX\vecp}\geq\normone{\bP_{\setX^c}\vecp}$. This would then imply that the difference vector $\vecp$ needs to be at least 50\%-concentrated to the set $\setP=\setX$ (of cardinality \nx), i.e., we require that $\epsilon_\setP\leq0.5$. Defining $\vecq=\error'-\error$ and $\setQ=\supp(\error'-\error)$, and noting that $\supp(\error)=\setE$ and $\supp(\error')\subseteq\setE$, it follows that $\abs{\setQ}\leq n_e$. This leads to the following chain of inequalities:
\begin{align}
 \nx\,n_e&\geq \abs{\setP}\abs{\setQ} \nonumber\\
& \geq \frac{\pos{(1+\coha)(1-\epsilon_\setP)-\abs{\setP}\coha} \pos{1-\cohb\left(\abs{\setQ}-1\right)}}{\cohm^2} \label{eq:XunknownBP2} \\
 & \geq \frac{1}{2}\, \frac{\pos{1-\coha(2\nx-1)} \pos{1-\cohb(n_e-1)}}{\cohm^2} \label{eq:XunknownBP3}
\end{align}
where~\eqref{eq:XunknownBP2} follows from the uncertainty relation in~\fref{thm:uncertainty} applied to the difference vectors $\vecp$ and $\vecq$ (with $\epsilon_\setP\leq0.5$ since \vecp is at least 50\%-concentrated to \setP and $\epsilon_\setQ=0$  since \vecq is perfectly concentrated to \setQ) and~\eqref{eq:XunknownBP3} is a consequence of  $\abs{\setP}=\nx$ and $\abs{\setQ}\leq n_e$. Rewriting~\eqref{eq:XunknownBP3}, we obtain
\be
\label{eq:XunknownBP4}
2\nx n_e\geq\frac{\pos{1\!-\!\coha(2\nx\!-\!1)} \pos{1\!-\!\cohb(n_e\!-\!1)}}{\cohm^2} = f(2\nx, n_e). 
\ee
Since~\eqref{eq:XunknownBP4} contradicts the assumption in~\eqref{eq:P0uk_assump}, this proves that \inputvec is the unique solution of $(\textrm{BP},\setE)$ applied to $\vecz=\dicta\inputvec+\dictb\error$. 
%


\sloppy

\section{Proof of~\fref{thm:P1uk_equivalence} }
\label{app:P1_uk_equivalence_proof}
We first show that condition~\eqref{eq:P0uk_assump} ensures that the columns of $\dictb_\setE$ are linearly independent. Then, we establish that \mbox{$\normtwo{\bR_\setE\veca_\indexa}>0$} for $\allonotwo{\ell}{\inputdimA}$. Finally, we show that the unique solution of (\Pzero), BP, and OMP applied to $\hat\vecz=\bR_\setE\dicta\bDelta\hat{\inputvec}$ is given by $\hat{\inputvec}=\bDelta^{-1}\inputvec$.

\subsection{The columns of $\dictb_\setE$ are linearly independent}
Condition~\eqref{eq:P0uk_assump} can only be satisfied if \mbox{$\pos{1\!-\!\cohb(n_e\!-\!1)}>0$}, which implies that \mbox{$n_e<1+1/\cohb$}. It was shown in~\cite{donoho2002,gribonval2003,tropp2004} that for a dictionary \dictb with coherence \cohb no fewer than $1+1/\cohb$ columns of \dictb can be linearly dependent. Hence, the $n_e$ columns of $\dictb_\setE$ must be linearly independent.

\fussy

\subsection{$\normtwo{\bR_\setE\veca_\indexa}>0$ for $\allonotwo{\ell}{\inputdimA}$}
We have to verify that condition~\eqref{eq:P0uk_assump} implies $\normtwo{\bR_\setE\veca_\indexa}>0$ for $\allonotwo{\ell}{\inputdimA}$. Since~$\bR_\setE$ is a projector and, therefore, Hermitian and idempotent, it follows that
\ba
	\normtwo{\bR_\setE\veca_\indexa}^2 & =\veca_\indexa^H\bR_\setE\veca_\indexa\label{eq:Risprojector1} \\
	& =  \abs{\veca_\indexa^H\bR_\setE\veca_\indexa}\nonumber\\
& \geq 1 - \underbrace{\abs{\veca_\indexa^H\bB_\setE\left(\bB^H_\setE\bB_\setE\right)^{-1}\bB^H_\setE\veca_\indexa}}_{\triangleq\,C_1} \label{eq:projectedcolumnslower1}
\ea
where~\eqref{eq:Risprojector1} is a consequence of $\bR_\setE^H\bR_\setE=\bR_\setE$, and~\eqref{eq:projectedcolumnslower1} follows from the reverse triangle inequality and $\normtwo{\veca_\indexa}=1$, $\allonotwo{\ell}{\inputdimA}$. Next, we derive an upper bound on $C_1$ according to 
\begin{align}
C_1 & \leq \lambda_\text{max}\lefto(\!\left(\bB^H_\setE\bB_\setE\right)^{\!-1}\right)\normtwo{\bB^H_\setE\veca_\indexa}^2 \label{eq:rayleighritz1}\\
       & \leq \lambda^{-1}_\text{min}\!\left(\bB^H_\setE\bB_\setE\right) n_e\,\cohm^2  \label{eq:equivalence_bound2}
\end{align}
where~\eqref{eq:rayleighritz1} follows from the Rayleigh-Ritz theorem~\cite[Thm. 4.2.2]{hornjohnson}  and~\eqref{eq:equivalence_bound2} results from 
\begin{align*}
\normtwo{\bB_\setE^H\veca_\indexa}^2=\sum_{\indexc\in\setE}\,\abs{\vecb^H_\indexc\veca_\indexa}^2 \leq n_e\,\cohm^2.
\end{align*}
Next,  applying  Ger\v{s}gorin's disc theorem \cite[Theorem~6.1.1]{hornjohnson}, we arrive at
\begin{align}
\lambda_\text{min}\lefto(\bB^H_\setE\bB_\setE\right)\geq \pos{1-\cohb(n_e-1)} \label{eq:equivalence_bound3}.
\end{align}
Combining \fref{eq:projectedcolumnslower1}, \fref{eq:equivalence_bound2}, and \fref{eq:equivalence_bound3}
 leads to the following lower bound on $\normtwo{\bR_\setE\veca_\indexa}^2$:
\begin{align} \label{eq:projectedcolumnslower2}
 \normtwo{\bR_\setE\veca_\indexa}^2 \geq 1- \frac{n_e\,\cohm^2}{\pos{1-\cohb(n_e-1)}}.
\end{align}
Note that if condition~\eqref{eq:P0uk_assump} holds for\footnote{The case  $\nx=0$ is not interesting, as $\nx=0$ corresponds to $\inputvec=\mathbf{0}_\inputdimA$ and hence recovery of $\inputvec=\mathbf{0}_\inputdimA$ only could be guaranteed.} $\nx\geq1$, it follows that $n_e\,\cohm^2<\pos{1-\cohb(n_e-1)}$ and hence the RHS of~\eqref{eq:projectedcolumnslower2} is strictly positive.
This ensures that  $\bDelta$ defines a one-to-one mapping.
%
 %
We next show that, moreover, condition~\eqref{eq:P0uk_assump} ensures that for every vector $\inputvec'\in\complexset^\inputdimA$ satisfying $\normzero{\inputvec'}\leq2\nx$, $\dicta\inputvec'$ has a nonzero component that is orthogonal to $\setR(\dictb_\setE)$. 
%
%
 
 %

\subsection{Unique recovery through (\Pzero), BP, and OMP}
\label{app:effectivecoherence}
We now need to verify that (\Pzero), BP, and OMP (applied to $\hat\vecz=\bR_\setE\dicta\bDelta\hat\vecx$) recover the vector $\hat{\inputvec}=\bDelta^{\!-1}\inputvec$ provided that~\eqref{eq:P0uk_assump} is satisfied.
This will be accomplished by deriving an upper bound on the coherence $\mu(\bR_\setE \dicta\bDelta)$ of the modified dictionary  $\bR_\setE\dicta\bDelta$, which, via the well-known coherence-based recovery guarantee~\cite{donoho2002,gribonval2003,tropp2004}
\begin{align} \label{eq:modifiedDictRecovery}
\nx<\frac{1}{2}\Big(1+\coh( \bR_\setE \dicta\bDelta)^{-1}\Big)
\end{align}
leads to a recovery threshold guaranteeing perfect recovery of $\hat{\inputvec}$. This threshold is then shown to coincide with~\eqref{eq:P0uk_assump}. 
%
%
More specifically, the well-known sparsity threshold in~\eqref{eq:classicalthreshold} guarantees that the unique solution of (\Pzero) applied to $\hat\vecz=\bR_\setE\dicta\bDelta\hat\inputvec$ is given by $\hat\inputvec$, and, furthermore, that this unique solution can be obtained through BP and OMP if \fref{eq:modifiedDictRecovery} holds.
%
%
It is important to note that $\normzero{\hat{\inputvec}}=\normzero{\inputvec}=\nx$. With 
\be
	[\bDelta]_{\ell,\ell} = \frac{1}{\normtwo{\bR_\setE \cola_\ell}}, \quad \ell=1,\ldots,N_a
\een
we obtain 
\begin{align} \label{eq:equivalence_1}
\mu( \bR_\setE \dicta\bDelta) = \max_{\indexb,\indexa,\indexa\neq\indexb} \frac{\abs{\veca_\indexb^H\bR_\setE^H\bR_\setE\veca_\indexa}}{\normtwo{\bR_\setE\veca_\indexb}\normtwo{\bR_\setE\veca_\indexa}}.
\end{align}

\sloppy

Next, we upper-bound the RHS of~\fref{eq:equivalence_1} by upper-bounding its numerator and lower-bounding its denominator. 
For the numerator we have
\begin{align}
\abs{\veca_\indexb^H\bR_\setE^H\bR_\setE\veca_\indexa} & = \abs{\veca_\indexb^H\bR_\setE\veca_\indexa} \label{eq:effdic1} \\
 & \leq \abs{\veca_\indexb^H\veca_\indexa} +  \abs{\veca_\indexb^H\bB_\setE\bB^\dagger_\setE\veca_\indexa} \label{eq:effdic2}\\
 & \leq \coha + \underbrace{\abs{\veca_\indexb^H\bB_\setE\left(\bB^H_\setE\bB_\setE\right)^{-1}\bB^H_\setE\veca_\indexa}}_{\triangleq\, C_2} \label{eq:equivalence_bound1}
\end{align} 
where~\eqref{eq:effdic1} follows from $\bR_\setE^H\bR_\setE=\bR_\setE$,~\eqref{eq:effdic2} is obtained through the triangle inequality, and~\eqref{eq:equivalence_bound1} follows from \mbox{$\abs{\veca_\indexb^H\veca_\indexa}\leq\coha$}. Next, we derive an upper bound on $C_2$ according to 
\begin{align}
C_2 & \leq \normtwo{\bB^H_\setE\veca_\indexb}\normtwo{\left(\bB^H_\setE\bB_\setE\right)^{-1}\bB^H_\setE\veca_\indexa} \label{eq:effdic3} \\
       & \leq  \normtwo{\bB_\setE^H\veca_\indexb}\vecnorm{\left(\bB^H_\setE\bB_\setE\right)^{-1}}\normtwo{\bB^H_\setE\veca_\indexa} \label{eq:effdic4}
\end{align}
where~\eqref{eq:effdic3} follows from the Cauchy-Schwarz inequality and~\eqref{eq:effdic4} from the   Rayleigh-Ritz theorem~\cite[Thm. 4.2.2]{hornjohnson}. Defining $\indexc=\argmax_{\indexb}\normtwo{\bB_\setE^H\veca_\indexb}$, we further have
\begin{align}
C_2   & \leq \vecnorm{\left(\bB^H_\setE\bB_\setE\right)^{-1}}\normtwo{\bB_\setE^H\veca_\indexc}^2 \nonumber \\
&=\lambda_\text{max}\lefto(\left(\bB^H_\setE\bB_\setE\right)^{-1}\right )\normtwo{\bB_\setE^H\veca_\indexc}^2. \nonumber
\end{align}
We obtain an upper bound on $C_2$ using the same steps that were used to bound $C_1$ in~\eqref{eq:rayleighritz1} --~\eqref{eq:equivalence_bound3}:
\be\label{eq:C2bound}
	C_2\leq  \frac{n_e\,\cohm^2}{C_b}
\ee
where $C_b=\pos{1-\cohb(n_e-1)}$.
Combining \fref{eq:equivalence_bound1} and \fref{eq:C2bound}
 leads to the following upper bound 
\begin{align} \label{eq:equivalence_bound3a}
 \abs{\veca_\indexb^H\bR_\setE^H\bR_\setE\veca_\indexa} \leq \coha + \frac{n_e\,\cohm^2}{C_b}.
\end{align}

\fussy

Next, we derive a lower bound on the denominator on the RHS of~\eqref{eq:equivalence_1}. 
To this end, we set $\indexd=\argmin_{\indexb}\normtwo{\bR_\setE\veca_\indexb}$ and note that
\begin{align}
\normtwo{\bR_\setE\veca_\indexb}\normtwo{\bR_\setE\veca_\indexa}& \geq\normtwo{\bR_\setE\cola_\indexd}^2\notag\\
&\geq 1 - \frac{n_e\,\cohm^2}{C_b} \label{eq:equivalence_bound5a}
\end{align}
where~\eqref{eq:equivalence_bound5a} follows from~\eqref{eq:projectedcolumnslower2}.
Finally, combining~\eqref{eq:equivalence_bound3a} and~\eqref{eq:equivalence_bound5a} we arrive at 
\begin{align}
\mu(\bR_\setE \dicta\bDelta) \leq \frac{\coha C_b + n_e\,\cohm^2}{C_b-n_e\,\cohm^2}.\label{eq:effectivecoherencebound1}
\end{align}
Inserting~\eqref{eq:effectivecoherencebound1} into the recovery threshold in~\eqref{eq:modifiedDictRecovery}, we obtain the following threshold guaranteeing recovery of $\hat\inputvec$  from $\hat{\vecz}=\bR_\setE \dicta\bDelta\hat{\inputvec}$ through $(\Pzero)$, BP, and OMP:  
\begin{align} \label{eq:equivalence_bound6}
 \nx< \frac{1}{2} \left(\frac{C_b(1+\coha)}{\coha C_b + n_e\,\cohm^2}\right).
\end{align}
%
%
Since $2\nx n_e\,\cohm^2\geq0$, we can transform~\eqref{eq:equivalence_bound6} into 
\begin{align}
2\nx n_e\,\cohm^2 &< C_b\pos{1-\coha(2\nx-1)}\nonumber\\
& = \pos{1-\cohb(n_e\,-1)}\pos{1-\coha(2\nx-1)}.\label{eq:finalexpression}
\end{align}
Rearranging terms in~\eqref{eq:finalexpression} finally yields 
\begin{align*}
  2\nx n_e<f(2\nx,n_e)
\end{align*}
which proves that~\eqref{eq:P0uk_assump} guarantees recovery of  the vector $\hat{\inputvec}$ (and thus also of $\inputvec=\bDelta\hat{\inputvec}$)  through (\Pzero), BP, and OMP.

\sloppy

\section{Proof of~\fref{thm:P0_nothingknown}}\label{app:P1_bothuk_proof}
Assume that there exists an alternative vector $\inputvec'$ that satisfies $\dicta\inputvec'\in (\{\vecz\} +\bigcup_{\setE\in\mathscr{P}}\setR(\dictb_\setE))$ (with $\mathscr{P}=\wp_{n_e}(\{1,\ldots,\inputdimB\})$) with $\normzero{\inputvec'}\leq\nx$. This implies the existence of a vector $\vece'$ with $\normzero{\error'}\leq n_e$ such that
%
\be
	\dicta\inputvec+\dictb\error=\dicta\inputvec'+\dictb\error'
\een
and therefore
\begin{align*}
\dicta(\inputvec-\inputvec')=\dictb(\error'-\error).
\end{align*}
From $\normzero{\inputvec}=\nx$ and $\normzero{\inputvec'}\leq\nx$ it follows that $\normzero{\inputvec-\inputvec'}\leq2\nx$. Similarly, $\normzero{\error}=n_e$ and $\normzero{\error'}\leq n_e$
imply $\normzero{\error'-\error}\leq2 n_e$.
Defining $\vecp=\inputvec-\inputvec'$ and $\setP=\supp(\inputvec-\inputvec')$, and, similarly, $\vecq=\error'-\error$ and $\setQ=\supp(\error'-\error)$, we arrive at
\begin{align}
	4\nx n_e & \geq \normzero{\vecp}\normzero{\vecq} = \abs{\setP}\abs{\setQ}\nonumber\\
	 & \geq \frac{\pos{1\!-\!\coha(\abs{\setP}\!-\!1)} \pos{1\!-\!\cohb(\abs{\setQ}\!-\!1)}}{\cohm^2} \label{eq:bothunknownproof1}\\ 
 & \geq \frac{\pos{1\!-\!\coha(2\nx\!-\!1)} \pos{1\!-\!\cohb(2n_e\!-\!1)}}{\cohm^2}\!=\!f(2\nx,2 n_e) \label{eq:bothunknownproof2}
\end{align}
where~\eqref{eq:bothunknownproof1} follows from the uncertainty relation in~\fref{thm:uncertainty} applied to the difference vectors $\vecp$ and $\vecq$ (with $\epsilon_\setP=\epsilon_\setQ=0$ since both \vecp and \vecq are perfectly concentrated to \setP and \setQ, respectively) and~\eqref{eq:bothunknownproof2} is a consequence of $\abs{\setP}\leq2\nx$ and $\abs{\setQ}\leq 2n_e$. Obviously,~\eqref{eq:bothunknownproof2} is in contradiction to~\eqref{eq:P0_nothingknown_assump}, which concludes the proof. 

\fussy

\end{document}